\documentclass[aps,prb,superscriptaddress,twocolumn]{revtex4}
\usepackage{graphicx}
\usepackage{bm}
\usepackage{amssymb}
\newcommand{\jhu}
{Department of Physics and Astronomy, Johns Hopkins University, Baltimore,
Maryland 21218, USA}
\newcommand{\mpipks}
{Max-Planck-Institut f\"ur Physik komplexer Systeme, 01187 Dresden, Germany}

\begin{document}

\title{Partial order from disorder in a classical pyrochlore antiferromagnet}

\author{Gia-Wei Chern}
\affiliation{\jhu}
\author{R. Moessner}
\affiliation{\mpipks}
\author{O. Tchernyshyov}
\affiliation{\jhu}

\date{\today}

\begin{abstract}
We investigate theoretically the phase diagram of a classical
Heisenberg antiferromagnet on the pyrochlore lattice perturbed by a
weak second-neighbor interaction $J_2$. The huge ground state
degeneracy of the nearest-neighbor Heisenberg spins is lifted by
$J_2$ and a magnetically ordered ground state sets in upon
approaching zero temperature. We have found a new, partially ordered
phase with collinear spins at finite temperatures for a
ferromagnetic $J_2$. In addition to a large nematic order parameter,
this intermediate phase also exhibits a layered structure and a bond
order that breaks the sublattice symmetry. Thermodynamic phase
boundaries separating it from the fully disordered and magnetically
ordered states scale as $1.87 J_2 S^2$ and $0.26 J_2 S^2$ in the
limit of small $J_2$.  The phase transitions are discontinuous. We
analytically examine the local stability of the collinear state and
obtain a boundary $T\sim J_2^2/J_1$ in agreement with Monte Carlo
simulations.
\end{abstract}

\maketitle

\section{Introduction}

Magnets with geometrical frustration\cite{moessner:06pt} have
received much attention as models of strongly interacting electronic
systems with unusual ground states, thermodynamic phases, and
excitations. The hallmark of strong frustration is a conspicuously
large degeneracy of the classical ground state: essentially, a
finite {\em fraction} of the degrees of freedom remains
unconstrained to the lowest temperatures. For discrete spins, this
manifests itself in the number of ground states scaling
exponentially with the system volume and thus giving rise to a
nonzero entropy density at absolute zero temperature.  Well-known
examples of that are the Ising antiferromagnet on the triangular
lattice\cite{wannier:1950, houtappel:1950} and spin
ice.\cite{bramwell:science} For continuous spins --- most saliently
for the Heisenberg antiferromagnet on the pyrochlore lattice --- the
classical ground states form a \textit{manifold} whose dimension is
proportional to the system volume.\cite{Moessner98PRB} In
that particular case, the classical model exhibits strong
short-range spin correlations but fails to exhibit any form of
conventional magnetic order down to the lowest temperatures
accessible in Monte Carlo simulations. The strong correlation
between the local motions of spins in this liquid-like phase
manifests itself as an emergent gauge structure in the low-temperature limit
and results in a dipolar form of the asymptotic spin correlations at
large separations. \cite{isakov:2004prl, henley:2005prb}

At the same time, the large degeneracy of the ground state makes
this system susceptible to all kinds of perturbations, which
certainly exist in real compounds.  For instance, the spin-lattice
coupling, arising from the dependence of exchange strength on the
atomic displacements,\cite{Kittel60} lifts the degeneracy through a
spin analog of the Jahn-Teller effect\cite{Tch02PRB} observed in
spinels ZnCr$_2$O$_4$\cite{PhysRevLett.84.3718} and
CdCr$_2$O$_4$.\cite{chung:247204}

\begin{figure}
\includegraphics[width=0.8\columnwidth]{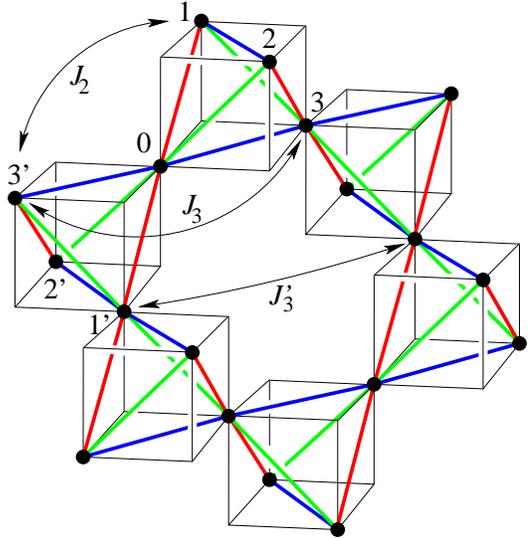}
\caption{Second and third-neighbor pairs on the pyrochlore lattice.
Since exchange paths giving rise to $J_3$ and $J_3'$ are inequivalent,
the two couplings may be different.  Numbers from 0 to 3 label the four
fcc sublattices.}
\label{fig:pyrochlore}
\end{figure}

\begin{figure}
\includegraphics[width=0.8\columnwidth]{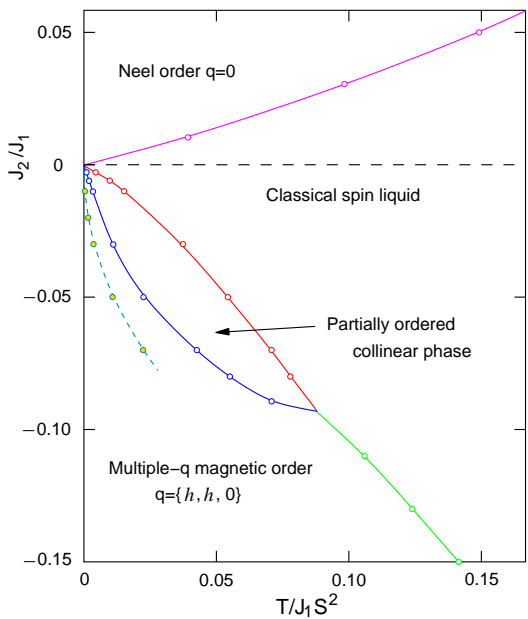}
\caption{Phase diagram of the model with antiferromagnetic first and
weak second-neighbor exchange of either sign on the pyrochlore
lattice. Open circles are numerically determined locations of
thermodynamic phase transitions (all first order); filled circles denote the
stability boundary of the collinear phase.  Solid lines are interpolated
phase boundaries; the dashed line is a boundary of local stability of
the collinear phase.  The wavenumber of the incommensurate magnetic
phase is $h \approx 3/4$.} \label{fig:phase-dgm}
\end{figure}

This naturally leads one to ponder the following questions. Can the
interplay of a weak perturbation with strong frustration lead to
interesting ordered phases? Are there any (intermediate) partially
ordered phases? What is the nature of the phase transitions between
such phases?  In this paper we discuss these questions in the
context of a classical Heisenberg antiferromagnet on the pyrochlore
lattice with interactions going beyond nearest neighbors.  Following
previous work by Reimers \textit{et al.}\cite{reimers:1991prb} and
by Tsuneishi \textit{et al.},\cite{tsuneishi:2007jpcm} we consider
the classical Heisenberg antiferromagnet on the pyrochlore lattice
with the Hamiltonian
\begin{equation}
    \mathcal{H} = J_1\sum_{\langle ij\rangle}
    \mathbf S_i\cdot\mathbf S_j
    + J_2\sum_{\langle\langle ij\rangle\rangle}
    \mathbf S_i\cdot\mathbf S_j,
    \label{eq-H0}
\end{equation}
where $\langle ij\rangle$ and $\langle\langle ij\rangle\rangle$
indicate pairs of first and second neighbors, respectively.  Given
the short-range nature of exchange forces, we work in the limit $J_2
\ll J_1$.  It is reasonable to expect that the influence of $J_2$
becomes noticeable only at low temperatures of order $J_2 S^2$, when
the system is already in the strongly correlated paramagnetic state,
in which it is constrained to fluctuate around the ground states of
the nearest neighbor exchange.  Using a combination of Monte Carlo
simulations and analytical arguments, we have mapped out the phase
diagram in the $J_2$--$T$ plane shown in Fig.~\ref{fig:phase-dgm}.

Antiferromagnetic second-neighbor exchange, $J_2>0$, significantly
reduces the frustration by selecting states in which spins within
any of the four fcc sublattices, comprising the pyrochlore lattice,
are parallel to one another.  We find a collinearly ordered phase of
the type $\langle \mathbf S_0 \rangle = -\langle \mathbf S_1 \rangle
= -\langle \mathbf S_2 \rangle =  \langle \mathbf S_3 \rangle$,
where the subscripts enumerate the fcc sublattices
[Fig.~\ref{fig:pyrochlore} and Fig. \ref{fig-afj2}(a)].  The
transition between the paramagnetic and antiferromagnetic phases is
discontinuous.

Ferromagnetic second-neighbor exchange, $J_2<0$, leaves the system
strongly frustrated.  A mean-field calculation by Reimers \textit{et
al.}\cite{reimers:1991prb} predicted a ground state with
incommensurate magnetic order.  While Tsuneishi \textit{et
al.}\cite{tsuneishi:2007jpcm} indeed observed Bragg peaks in the
spin structure factor obtained through a Monte Carlo simulation for
$J_2=-0.1 J_1$, they also noted that the spins remained dynamic,
failing to freeze.  We show that the observed locations of the Bragg
peaks are compatible with the results of Reimers \textit{et al.}, so
that the low-temperature phase is most likely magnetically ordered.

The main focus of our paper is a peculiar \textit{partially ordered}
phase sandwiched between the paramagnet and the magnetically ordered
state for weak enough ferromagnetic $J_2$, namely $-0.09 J_1
\lesssim J_2 < 0$. In the intermediate phase, the spins display
collinear order; furthermore, they exhibit magnetic order within a
thin $\{100\}$ layer but no order across different layers.  The
partial order can be characterized by a combination of a director
$\hat \mathbf n$ specifying a global spin axis, a Potts ($Z_3$)
variable $q = (100)$, (010), or (001) specifying the direction of
the layers, and an Ising ($Z_2$) variable $\sigma_n$ {\em for each
layer} identifying one of the two possible spin orientations within
a layer.  The order is partial in the sense that the Ising variables
$\{\sigma_n\}$ randomly pick values of $+1$ and $-1$ with no
discernible correlations between adjacent layers.  The partially
ordered state is bounded by first-order transitions on both the high
and low-temperature sides.

Similar partial order has been previously found in a $1/S$ treatment
of the Heisenberg antiferromagnet on the checkerboard lattice, also
known as the square lattice with crossings, a two-dimensional analog
of the pyrochlore.\cite{ot:2003prb}  In both systems, the distinct
layered states are \textit{not} related to one another by a symmetry
of the Hamiltonian and simply arise as different local minima of the
free energy.  Free-energy barriers separating them may be large enough in
practice for the system not to be ergodic and instead to remain in one of these
minima forever.

Since the energy of the partially ordered collinear state is greater
than that of the low-temperature multiple-$\mathbf q$ magnetic
order, entropic selection plays a crucial role in the stabilization
of the intermediate phase. This is consistent with the general
observation that states with collinear spins tend to have softer
thermal fluctuations and therefore have a lower free energy at
finite temperatures.\cite{Moessner98PRB,henley:89} A similar
collinear phase has been reported in the Monte Carlo study of a
$J$-$J'$ model which interpolates between the pyrochlore and the fcc
lattices. \cite{pinettes:02}

While we have focused on the role of second-neighbor exchange $J_2$
in the formation of magnetic order on the pyrochlore lattice, our
results also shed light on the role of third-neighbor interactions
$J_3$ (see Fig.~\ref{fig:pyrochlore}).  In view of strong
correlations between nearest-neighbor spins developing at
temperatures well below $J_1 S^2$, the properties of the system
depend not on $J_2$ and $J_3$ separately but on their linear
combination $J_2 - J_3$. Indeed, the relative shift in energy for
any pair of ground states of the nearest-neighbor exchange due to  a
small $J_3$ is identical to the effect of a $J_2$ of the same
magnitude and opposite sign. Thus our findings should also be of
relevance for the more general case of a pyrochlore antiferromagnet
with small $J_2$ and $J_3$.

The remainder of this paper is organized as follows.  In
Sec.~\ref{sec:low-T} we briefly discuss the nature of magnetically
ordered phases at low temperatures for both signs of the
second-neighbor coupling $J_2$.  Sec.~\ref{sec:int-T} presents the
main subject of this work, the partially ordered phase found at
intermediate temperatures on the ferromagnetic side of $J_2$.
Stability of the partially ordered state and its phase boundaries
are examined in Sec.~\ref{sec:stability}.  We conclude with a
discussion of these results in Sec.~\ref{sec:discussion}.

\section{Low-temperature ordered phases}
\label{sec:low-T}

Since the phase transitions shown in Fig. \ref{fig:phase-dgm} are
strongly discontinuous and occur at very low temperatures, the
metastable states close to the coexisting region are rather
long-lived. Conventional histogram methods with local Metropolis
updates are ineffective in determining the critical points due to a
large energy barrier separating the metastable state from the true
ground state. Instead, we settled on using a method proposed by M.
Creutz \textit{et al.}, \cite{creutz:79} in which a mixed phase with
the two coexisting states each occupying half the lattice is
constructed first. By thermalizing the mixed phase at various
temperatures, the critical point is determined when neither of the
two states prevail the system during the relaxation process. Since
the multiple-$\mathbf q$ magnetic order has an extended unit cell
with a period of about 4 cubic lattice constants, systems used in
our mixed-phase simulations contain $8^3$ cubic unit cells, with a
total spin $N = 16\times 8^3$.

\subsection{Antiferromagnetic $J_2$: low frustration}

\begin{figure}
\includegraphics[width = 0.30\columnwidth]{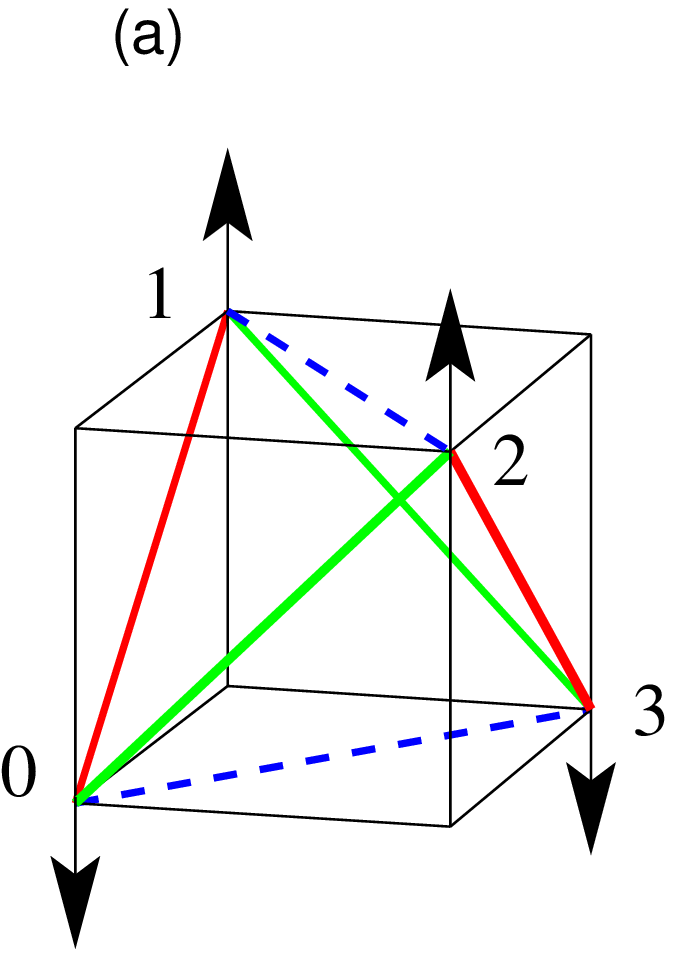}
\includegraphics[width = 0.68\columnwidth]{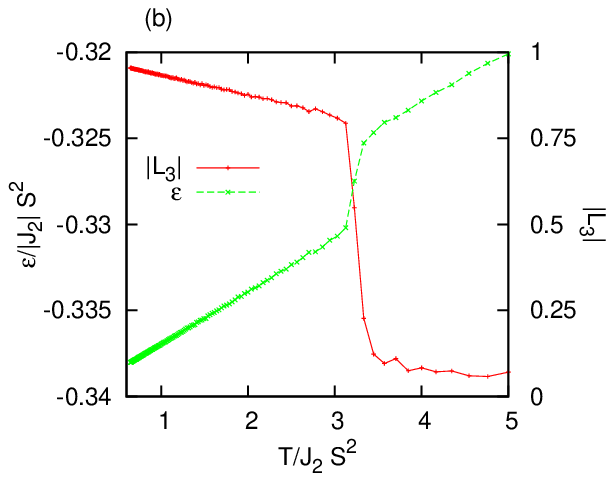}
\caption{\label{fig-afj2} (a)A ${\bf q} = 0$ N\'eel order for model
with an antiferromagnetic $J_2$ (ferromagnetic $J_3$). The order
parameter is one of the three staggered magnetization ${\bf L}_3 =
({\bf S}_0+{\bf S}_1 - {\bf S}_2 - {\bf S}_3)/4
S$.\cite{chern:060405} (b) The phase transition between the
paramagnetic and antiferromagnetic phases for $J_2= 0.01 J_1$. The
simulated system has $16\times 8^3$ spins. The energy density
$\varepsilon = (E-E_0)/6 N_s$, where $E_0 = -N_s J_1$ is the ground
state energy of nearest-neighbor interactions.}
\end{figure}

In the limit $J_2 \ll J_1$, magnetic ordering takes place at a
temperature $T_c = \mathcal O(J_2 S^2)$. The nature of this ordering
is best understood by appealing to the fact that a weak
third-neighbor coupling $J_3 \ll J_1$ (Fig.~\ref{fig:pyrochlore})
selects among the nearest-neighbour ground states in the same way as
a second-neighbor coupling $J_2$ of the same strength and opposite
sign, as explained in Appendix \ref{app:equiv}. (We here note in
passing that, since the strength of coupling depends on the exchange
paths and not the interatomic distance alone, sometimes $J_3$ may be
as big as $J_2$.  For instance, \textit{ab initio} calculations show
that in CdCr$_2$O$_4$ $J_3$ exceeds $J_2$ in
magnitude.\cite{chern:060405, yaresko:2008}) This insight is useful
as the resulting ordered pattern can be understood in a more
straightforward way by analyzing the effect of $J_3$.  To see that,
note that the pyrochlore lattice consists of four fcc sublattices
and that third neighbors on the pyrochlore lattice belong to the
same fcc sublattice (Fig.~\ref{fig:pyrochlore}).  Thus a
ferromagnetic exchange $J_3<0$ is not frustrated and will be
absolutely minimized by a state where spins within the same fcc
sublattice are parallel to one another.

A translationally invariant four-sublattice ground state was
predicted for the pyrochlore antiferromagnet with a ferromagnetic
$J_3$ by Reimers \textit{et al.}\cite{reimers:1991prb}  The same can
be expected for an antiferromagnetic second-neighbor coupling
$J_2>0$. In both cases the energy of the further-neighbor exchange
is minimized by a ferromagnetic order $\langle \mathbf S_i \rangle$
within the individual sublattices. Consequently any configuration
satisfying $\sum_{i=0}^3 \langle \mathbf S_i \rangle = 0$ is a
ground state at the mean-field level. Thermal fluctuations
nonetheless favor those with collinear spins. \cite{Moessner98PRB}
This is indeed what we obtained in the Monte Carlo simulations
(Fig.~\ref{fig-afj2}): a $\mathbf q=0$ N\'eel state with an
up-up-down-down spin configuration on every tetrahedron is found to
be the ground state for an antiferromagnetic $J_2$. This collinear
magnetic state is separated by a discontinuous transition line from
the high-temperature cooperative paramagnetic state. As shown in
Fig. \ref{fig-afj2}(b), both the energy density $\varepsilon$ and
the staggered magnetization $\mathbf L_3 = (\mathbf S_0 - \mathbf
S_1 - \mathbf S_2 + \mathbf S_3)/4S$ show a clear jump at the
transition temperature $T_c \approx 3.2\, J_2 S^2$.

\subsection{Ferromagnetic $J_2$: high frustration}
\label{sec-high-frustration}

The case of a ferromagnetic second-neighbor coupling, $J_2<0$, is
similar to that of $J_3>0$.  An antiferromagnetic coupling on an fcc
lattice is frustrated, so that this time one may expect a more complex
magnetic order.  Indeed, Reimers's mean-field calculation yields an
incommensurate magnetic order with a wavevector $\mathbf q=2\pi(h,h,0)$
in the case of a ferromagnetic $J_2$.

\begin{figure}
\includegraphics[scale = 0.32]{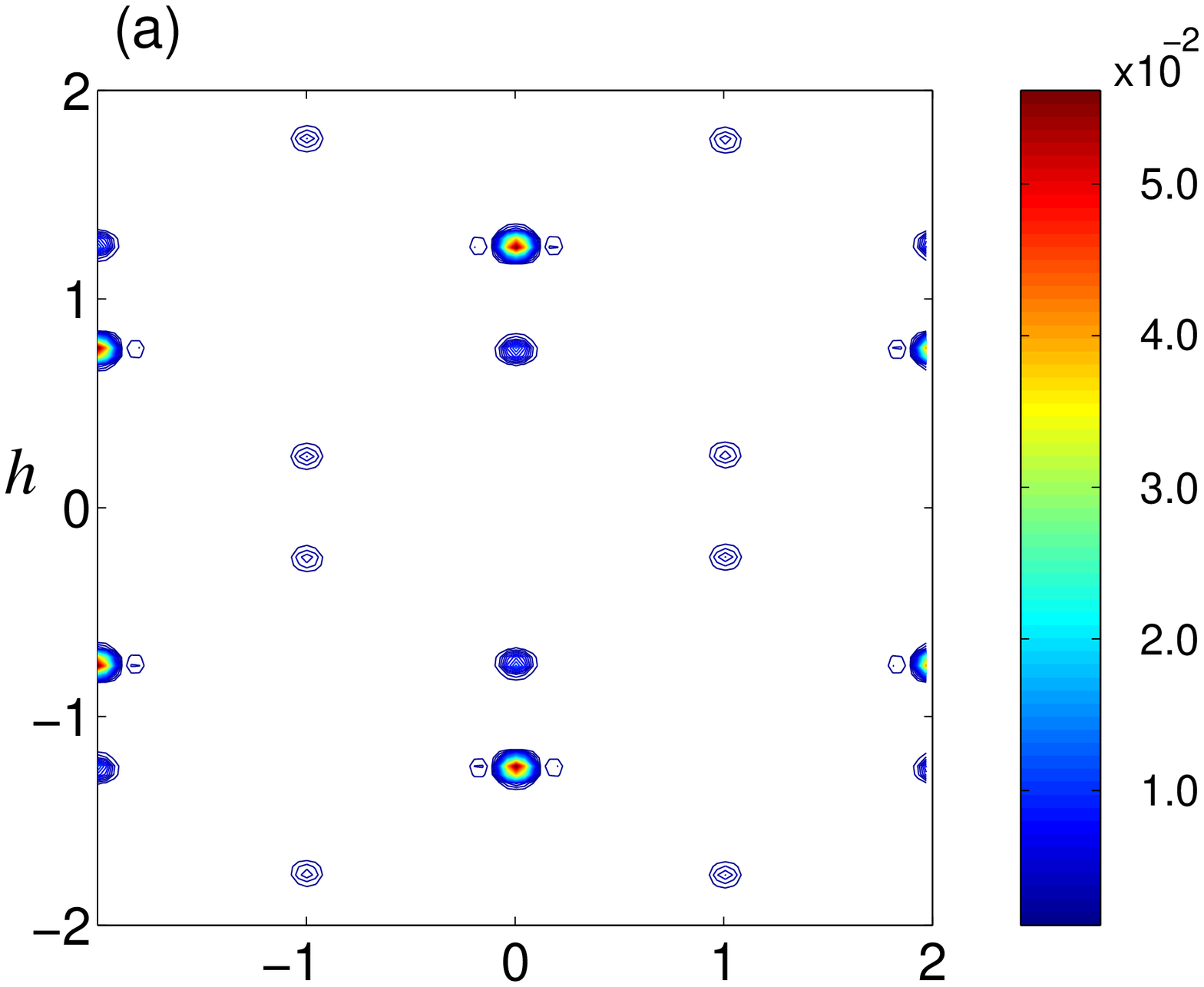}
\includegraphics[scale = 0.32]{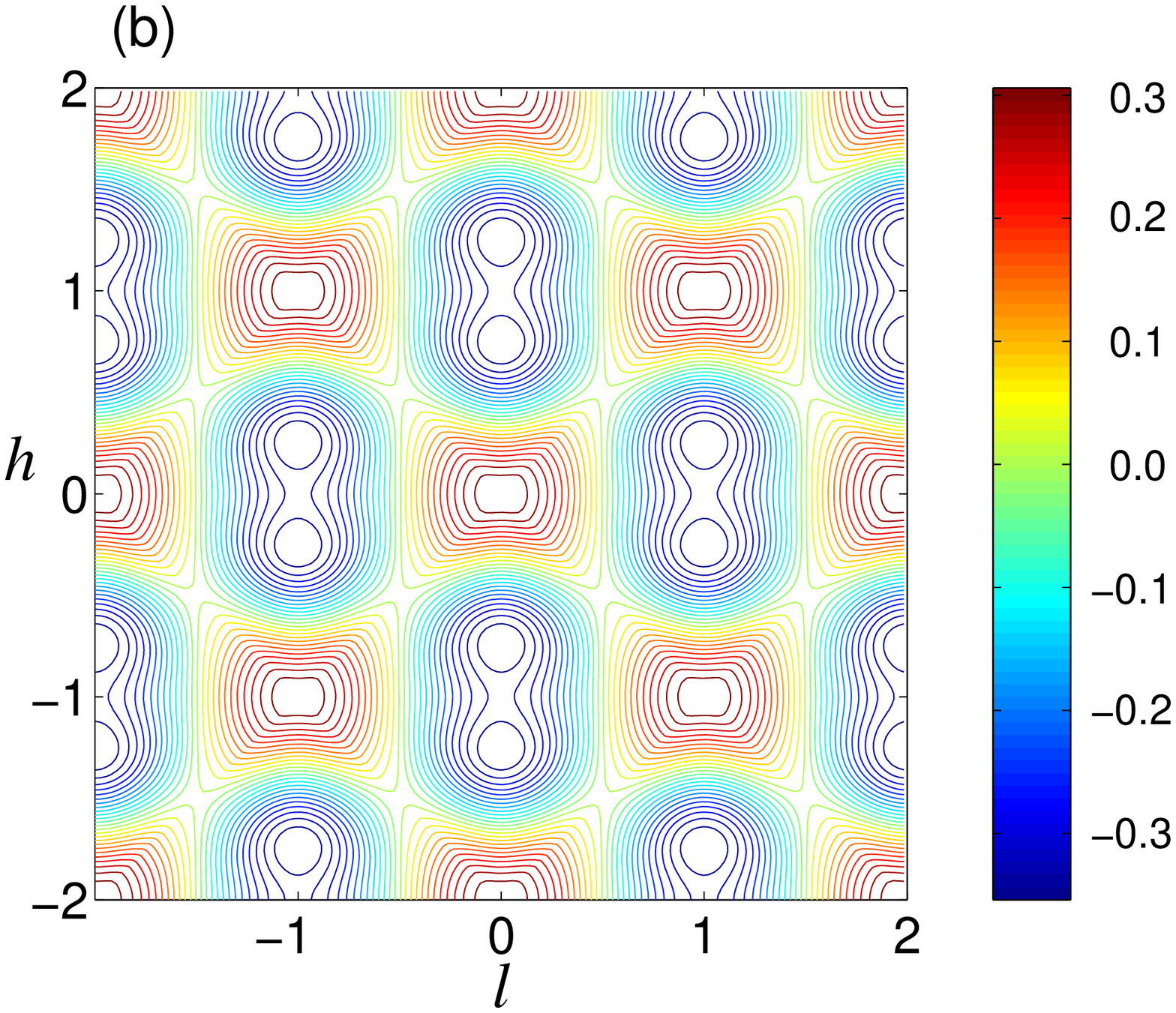}
\caption{\label{fig-fq} (a) The spin structure factor of the
low-temperature ordered state at wavevectors $\mathbf q =
2\pi(h,h,l)$. The state was obtained from a Monte-Carlo simulation
with a system with $16\times 8^3$ spins for a ferromagnetic $J_2$;
the temperature was $T = 0.2\,|J_2|$. (b) Minimum eigenvalue of the
exchange matrix $J_{mn}(\mathbf q)$ at wavevectors $\mathbf q =
2\pi(h,h,l)$ for $J_2=-J_1/10$. The satellite peaks at ${\bf
q}\approx 2\pi(\frac{5}{4},\frac{5}{4},\pm 0.1)$ might be due to the
finite-size effect for an incommensurate spin order.}
\end{figure}

We have performed Monte Carlo simulations on the pyrochlore lattice with
periodic boundary conditions measuring 8 cubic unit cells in each
direction. The simulations were done for $J_2 = -0.1\,J_1$. They
revealed a state with magnetic Bragg peaks at incommensurate lattice
momenta near $2\pi\{3/4,\,3/4, 0\}$  and other equivalent positions.
Fig.~\ref{fig-fq}(a) shows two inequivalent Bragg peaks, $\mathbf q
\approx 2\pi(3/4,\,3/4,0)$ and $-2\pi(3/4,\,3/4,0)$, the rest being
related to these two by a reciprocal lattice vector. Bragg peaks
with comparable intensities are found at other wavevectors related
to the above two by point-group symmetries. This multiple-$\mathbf
q$ N\'eel order is consistent with the ground states of
(\ref{eq-H0}) in the spherical approximation, in which the local
length constraints $|\mathbf S_i|=S$ are replaced by a global one
$\sum_{i=1}^N\,|\mathbf S_i|^2 = N S^2$. Introducing the Fourier
transform $\mathbf S_i = \sum_{\bf q} \mathbf S_m({\bf q})
e^{i\mathbf q\cdot\mathbf r_i}$ [the site index $i = (m,\mathbf
r_i)$, where $m$ is the sublattice index], the exchange interaction
(\ref{eq-H0}) becomes
\begin{equation}
    \label{eq-H1}
    \mathcal{H} = \frac{N}{4}\sum_{{\bf q}}
    \sum_{m,n=0}^3 J_{mn}({\bf q})
    \,{\bf S}_m({\bf q})\cdot{\bf S}_n(-{\bf q}).
\end{equation}
The Fourier components $\mathbf S_m(\mathbf q)$ are subject only to
a global constraint $\sum_{m,\bf q} |\mathbf S_m({\bf q})|^2 = S^2$.
The matrix $J_{mn}(\mathbf q)$ is the Fourier transform of the
exchange interaction $J_{ij}=J_{mn}({\bf r}_i - {\bf r}_j)$. Its
explicit form with interactions up to the fourth nearest neighbors
can be found in Ref. \onlinecite{reimers:1991prb}.

Expanding $\mathbf S_m(\mathbf q) =\sum_a U^a_{\mathbf q,\,m}\,\bm
\Phi^a_{\mathbf q}$ in terms of the eigenvectors $U^a_{\mathbf q,
m}$ of the exchange matrix $J_{mn}(\mathbf q)$ yields the energy as
a function of the expansion coefficients $\bm \Phi^a_{\mathbf q}$:
\begin{equation}
    E = \frac{N}{4} \sum_{\mathbf q}\sum_{a=1}^4
    \lambda^a_{\mathbf q} |\bm \Phi^a_{\mathbf q}|^2,
\end{equation}
where $\lambda^a_{\mathbf q}$ is the corresponding eigenvalue of
$J_{mn}(\mathbf q)$. With the normalization
$\sum_{m=0}^3|U^a_{\mathbf q,m}|^2=1$, the vectors $\bm
\Phi^a_{\mathbf q}$ satisfy $\sum_{\mathbf q} \sum_a
|\bm\Phi^a_{\mathbf q}|^2 = S^2$. The ground state energy of
(\ref{eq-H1}) is thus $E_0 = N S^2 \lambda_{\rm min}$, where
$\lambda_{\rm min}$ is the lowest eigenvalue $\lambda^a_{\mathbf
q}$.

For the nearest-neighbor interaction only, the two lowest
eigenvalues are $\mathbf q$-independent, $\lambda^1_{\mathbf q}
=\lambda^2_{\mathbf q} = -J_1$, reflecting the degenerate nature of
the magnetically ordered ground state. This degeneracy is lifted by
the introduction of $J_2$ as discussed by Reimers {\em et al.}
\cite{reimers:1991prb}  A contour plot of the lowest eigenvalue of
the exchange matrix as a function of the wavevector $\mathbf
q=2\pi(h,h,l)$ for $J_2<0$ is shown in Fig. \ref{fig-fq}(b). It can
be seen from Fig. \ref{fig-fq} that the peaks of the spin structure
factor appear at the same locations as the minima of exchange
energy, namely at 12 incommensurate wavevectors $\mathbf q^* =
2\pi\{h^*,h^*,0\}$, where $h^*\approx 3/4$ depends weakly on the
ratio $J_2/J_1$.
For small $J_2/J_1$, 
$h^*=a_0+a_1 (J_2/J_1)+\mathcal{O}((J_2/J_1)^2)$, where
\begin{eqnarray}
    a_0 &=&\frac{1}{\pi}\arccos[(4\sqrt{3}-9)/3]=0.7427,\nonumber \\
    a_1 &=& \frac{44}{3\pi\sqrt{9654+5574\sqrt{3}}}=0.0336.
    \label{eq:h-j2}
\end{eqnarray}

\begin{figure}
\includegraphics[height=0.55\columnwidth]{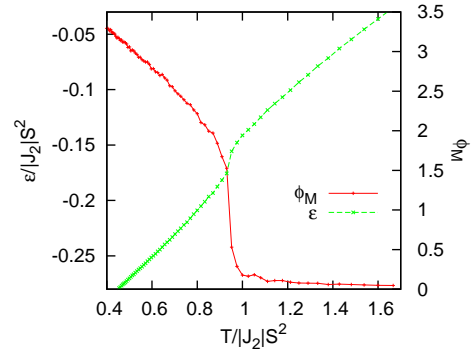}
\caption{\label{fig-trans-m} The phase transition between the
paramagnetic and antiferromagnetic phases for $J_2= -0.1 J_1$. The
simulated system has a total of $N=16\times 8^3$ spins. The
normalized energy density $\varepsilon = (E-E_0)/6 N$, where $E_0 =
-N J_1 S^2$ is the ground state energy of nearest-neighbor
interactions. $\phi_M$ is the second moment of the magnetic order
parameters.}
\end{figure}

The magnetic order is described by the order parameter composed of
12 vector amplitudes $\bm \Phi_{\mathbf q^*}$.\cite{reimers:1991prb}
A detailed characterization of this magnetic state is deferred to a
future publication. Fig.~\ref{fig-trans-m} shows the temperature
dependence of the energy density $\varepsilon$ and the magnitude of
the order parameters $\phi_M = \sum_{\mathbf q^*} |\bm\Phi_{\mathbf
q^*}|^2$. Both exhibit a clear jump at $T_c \approx 0.95|J_2| S^2$,
indicating a first-order transition. This is also confirmed by a
double-peak structure in the energy histogram at the transition
temperature. Similar results were obtained for $J_2 \lesssim
-0.09\,J_1 $ where the magnetic phase is separated from the
high-temperature spin liquid phase by a first-order phase transition
as indicated in Fig.~\ref{fig:phase-dgm}.

\section{Partially ordered phase}
\label{sec:int-T}

As discussed in the Introduction, an intermediate phase with
collinear spins exists at finite temperatures for a small
ferromagnetic coupling $J_2<0$. The appearance of collinearity is
not totally unexpected as it is well known that collinear states are
in general favored by thermal fluctuations in magnets with
frustrated exchange interactions. \cite{Henley89PRL} The fact that
the system remains frustrated even in the presence of a
ferromagnetic $J_2$ makes the existence of the nematic phase
possible. From another perspective, the classical nearest-neighbor
Heisenberg spins on the pyrochlore lattice evade the thermal
selection only marginally.\cite{Moessner98PRB} The introduction of a
ferromagnetic $J_2$ reduces the dimension of ground-state manifold,
thus permitting thermal fluctuations to stabilize collinear states.

\begin{figure}
\includegraphics[height=0.52\columnwidth]{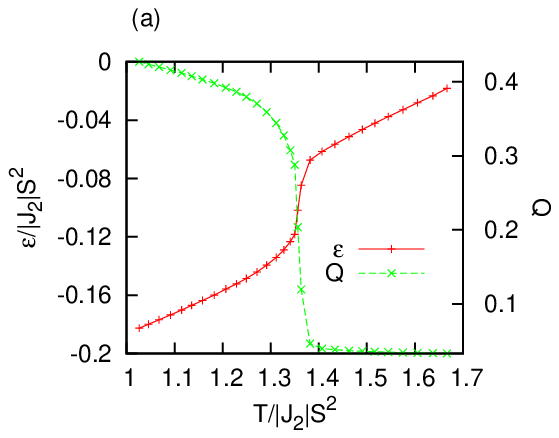}
\includegraphics[height=0.52\columnwidth]{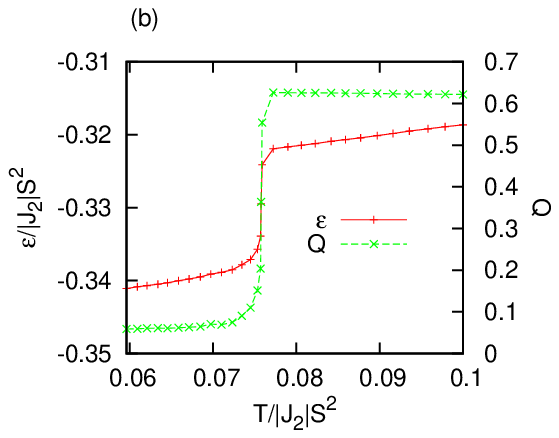}
\caption{\label{fig-transition} Transitions between (a) the
paramagnetic and nematic phases, and (b) the nematic and N\'eel
phases, for $J_2=-0.01 J_1$. A parallel-tempering Monte Carlo method
was employed to simulate a system with $16\times 4^3$ spins. The
normalized energy density $\varepsilon = (E-E_0) /6 N$, where $E_0 =
-N J_1 S^2$ is the ground state energy of nearest-neighbor
interactions. $Q$ is the spin nematic order parameter.}
\end{figure}

\subsection{Nematic order}

To demonstrate that spins indeed become collinear in the
intermediate phase, we have obtained from Monte Carlo simulations
the nematic order parameter $Q$ defined as the largest eigenvalue of
the traceless tensor $Q_{\mu\nu} = \langle S_{\mu} S_{\nu}/S^2 -
\delta_{\mu\nu}/3\rangle$, \cite{Chaikin} where $S_\mu$ represents
Cartesian components of a spin.  It vanishes in a totally disordered
state and attains the maximal value of 2/3 for parallel spins.

The thermodynamic behavior of the system with $J_2 = -0.01\,J_1$ in
the vicinity of the phase transitions is illustrated in
Fig.~\ref{fig-transition}. The simulation was done on the pyrochlore
lattice with periodic boundary conditions measuring 4 cubic unit
cells in each direction, giving a total of $N = 16 \times 4^3 =
1024$ spins. To improve the equilibration process, we employed
parallel tempering\cite{hukushima:1996jpsj, trebst:2004pre} with 30
replicas. The energy density $\varepsilon$ and the nematic order
parameter $Q$ are  shown as functions of temperature near $T_{c1}$
[paramagnet to partially ordered phase, Fig.~\ref{fig-transition}
(a)] and $T_{c2}$ [partially ordered phase to antiferromagnet, Fig.
\ref{fig-transition} (b)]. The energy density shows a clear
discontinuity at both transitions.   Extrapolating the energy curve
from the partially ordered phase to $T=0$ yields a density
$\varepsilon_L = -|J_2|/3$ characteristic of a layered state to be
discussed below. Likewise, the order parameter $Q$ extrapolates to
the maximal attainable value of 2/3 characteristic of collinear
spins. Below $T_{c2}$, the antiferromagnetic state seems to have a
residual nematic order with $Q\approx 0.05$, which may be intrinsic
to the low-temperature ordered state, or a finite-size effect.

\subsection{Bond order}

\begin{figure}
\includegraphics[width=0.92\columnwidth]{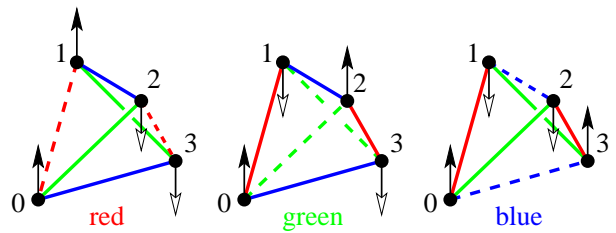}
\caption{The three distinct collinear states of a tetrahedron.
Frustrated bonds (with parallel spins) are shown as dashed lines.}
\label{fig:rgb}
\end{figure}

Nematic order alone does not provide a full characterization of this
phase: four spins on a tetrahedron have three distinct collinear
states not related to each other by a global rotation of the spins.
They are labeled red, green, and blue in Fig.~\ref{fig:rgb}. These
states differ from one another by the location of frustrated bonds
$\langle ij \rangle$ that involve parallel spins.  Since the global
direction of the spins is already captured by the nematic order
parameter $Q_{\mu\nu}$, further characterization can be made by
using scalar quantities, such as bond variables $f_{ij} \equiv
\langle \mathbf S_i \cdot \mathbf S_j \rangle$.  At temperatures
well below $J_1 S^2$ only two (out of six) bond variables of a
tetrahedron are independent:\cite{Tch02PRB}
\begin{eqnarray}
f_1 &=& \frac{f_{01} + f_{23} + f_{02} + f_{13} - 2 f_{03} - 2 f_{12}}
             {\sqrt{12}},
\nonumber\\
f_2 &=& \frac{f_{01} + f_{23} - f_{02} - f_{13}}{2}.
\label{eq:f}
\end{eqnarray}
The vector $\mathbf f = (f_1, f_2)$ takes on values in a triangular
domain with the three collinear states in its corners.

What kind of bond order might one expect in the intermediate phase?
To answer this question, let us again use the equivalence between a
ferromagnetic $J_2$ and an antiferromagnetic $J_3$.  The latter
promotes antiparallel orientations for spins 3 and $3'$
(Fig.~\ref{fig:pyrochlore}), which means---for a collinear state of
spins---that one of the bonds 03 and $03'$ is frustrated and the
other is satisfied. (Bergman \textit{et al.}\cite{bergman:134409}
showed that such states---satisfying the ``bending rule" for
frustrated bonds in zero applied field---are also favored by quantum
fluctuations of spins.) In other words, adjacent tetrahedra will be
in states of different color.  This is reminiscent of the
antiferromagnetic Potts model with 3 states: red, green and blue in
Fig.~\ref{fig:rgb}. A collinear state of the pyrochlore
antiferromagnet is fully specified by the global spin director and
the colors of all tetrahedra.  Note however that colors of
tetrahedra are not completely independent: the number of satisfied
bonds ($\mathbf S_i \cdot \mathbf S_j = -S^2$) must be even along
any closed loop.  Nonetheless, the parameterization in terms of Potts
variables serves a useful purpose.  One of the phases of the
antiferromagnetic Potts model on a bipartite lattice has a broken
sublattice symmetry (BSS): one sublattice is dominated by one color,
while the other is randomly populated by the two remaining
colors.\cite{grest:1981prl, lapinskas:1998prl}  With this state in
mind, we have measured the average bond variables in the
intermediate phase in the Monte Carlo simulations.

\begin{figure}
\includegraphics[scale=0.95]{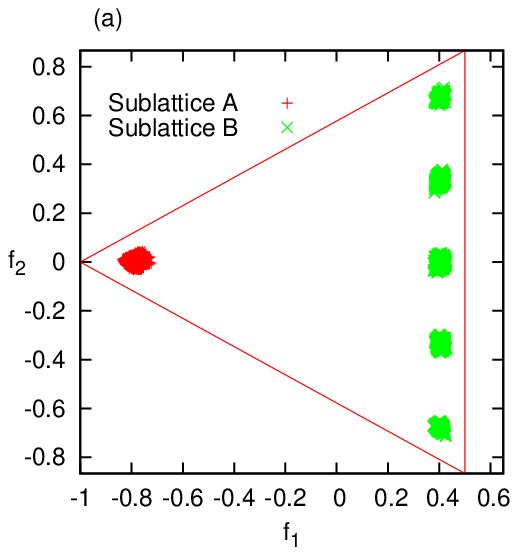}
\includegraphics[scale=0.95]{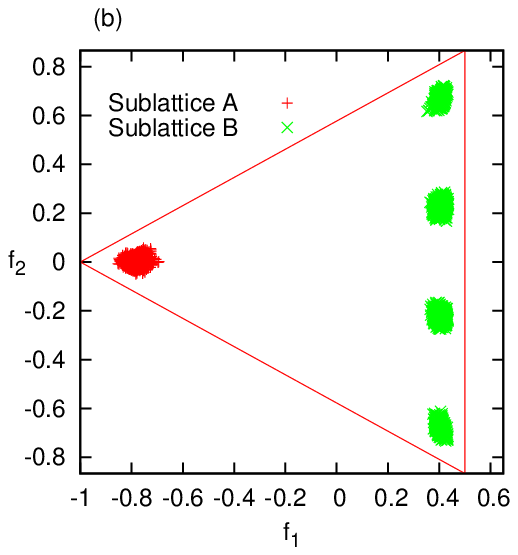}
\caption{\label{fig-fAB} The distribution of bond vector of the two
sublattices $\mathbf f_A$ and $\mathbf f_B$ in the nematic phase.
The simulated system has (a) 8 layers and (b) 6 layers of tetrahedra
in one sublattice. The bond vector $\mathbf f$ has been normalized
such that the three collinear states, blue, red, and green, are at
vertices $(-1,0)$, $(\frac{1}{2}, \frac{\sqrt{3}}{2})$, and
$(\frac{1}{2}, \frac{-\sqrt{3}}{2})$, respectively.}
\end{figure}

The Monte Carlo averages of the bond doublet (\ref{eq:f}) for
sublattices $A$ and $B$ are shown in Fig.~\ref{fig-fAB}. The value
of $\mathbf f$ for sublattice $A$ is narrowly distributed in the
vicinity of the collinear blue state, indicating that all tetrahedra
of sublattice $A$ are in this state. There are no blue tetrahedra on
sublattice $B$, as one might expect from the analogy with the
antiferromagnetic Potts model.  For the BSS phase, where each site
is red or green with equal probabilities, one expects a continuous
distribution of $\mathbf f$ in the middle of the opposing edge of
the triangle connecting the green and red corners.  Instead, we find
that sublattice $B$ has discrete fractions of red tetrahedra: e.g.
0, 1/4, 1/2, 3/4, and 1 in a system with 8 layers of tetrahedra in
one sublattice [Fig. \ref{fig-fAB}(a)].

This discreteness is a finite-size effect: an examination of
individual microstates shows that the intermediate phase has a
layered structure for bond variables on sublattice $B$: tetrahedra
within the same layer in the $xy$ plane have the same color.  The
origin of the layered structure on one of the sublattices can be
traced to the same constraint on the colors around a closed loop.
See Appendix \ref{app:constraint} for details. For example, the
simulated system of Fig. \ref{fig-fAB}(a) contained 8 layers of
tetrahedra within a sublattice. If the layers could be colored red
and green independently of one another, one would expect to find the
fractions of either color proportional to 1/8.  However, periodic
boundary conditions create constraints on the number of satisfied
bonds in the direction perpendicular to the layers, so that each
lattice can only have an even number of layers of either color.
Hence the fractions proportional to 1/4. Similarly, for a system in
which each sublattice has 6 layers of tetrahedra, the fraction of
red layers is 0, 1/3, 2/3, and 1 [Fig. \ref{fig-fAB}(b)].

\begin{figure}
\includegraphics[width=0.74\columnwidth]{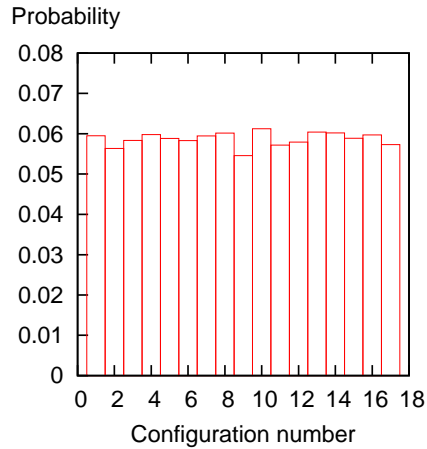}
\caption{\label{fig-hist} Histogram of 17 distinct collinear
layered structures obtained by replica-exchange Monte Carlo
simulation. The system has $16\times 4^3$ spins. The configuration
number labels 17 topologically distinct layered states subject
to the periodic boundary condition.}
\end{figure}

To verify this observation more directly, we performed a
replica-exchange Monte Carlo simulation on a system with $4^3$
conventional cubic cells. $16\times 4^3$ spins are divided into 8
layers of tetrahedra in each sublattice. A particular layered state
with collinear spins is described by a sequence of Ising variables
$\{\sigma_1,\sigma_2,\cdots\sigma_8\}$ (see Appendix
\ref{sec:layered-state}). With periodic boundary conditions, 17
distinct configurations are used in a replica-exchange Monte Carlo
simulation. The Ising sequences corresponding to these 17 layered
states are listed in Table I. In each exchange cycle, a fixed number
of Metropolis sweeps are performed on individual replicas of the
system, each of which corresponds to a particular layered state.
Then different replicas are exchanged according to detailed balance,
thus ensuring thermodynamic equilibrium. A histogram of the
occurrence of the 17 configurations in a chosen replica is shown in
Fig. \ref{fig-hist}. The almost equal probability of occurrence
implies a vanishing spin order after averaging over the different
configurations.

The layered structure of the intermediate phase spontaneously breaks
the rotational and translational symmetries of the pyrochlore
lattice. A collinear N\'eel order exists within an individual layer
of tetrahedra but not across the layers if the colors on one
sublattice are indeed random. At the mean-field level, the collinear
states in the partially ordered phase belong to a larger class of
(generally non-collinear) layered states with the same exchange
energy. A discussion of the general layered states is presented in
Appendix \ref{sec:layered-state}. As already mentioned  previously,
since collinear spins tend to have softer magnon spectrum, those
layered states with collinear spins are favored by thermal
fluctuations.

The two phase boundaries enclosing the intermediate phase are both
discontinuous transitions.  The critical temperatures determined by
the mixed-phase method \cite{creutz:79} are linear in $J_2$:
$T_{c1} \sim 1.87 |J_2|S^2$ and $T_{c2} \sim 0.26 |J_2|S^2$ as
$T \to 0$. Our numerical simulations seem to indicate
that the intermediate phase is globally stable in the temperature
regime $T_{c2} < T < T_{c1}$: in the mixed state, the collinear phase
gradually takes over the entire lattice.  We do not have analytical
arguments to back up the global stability of the intermediate collinear
phase: such an analysis would require knowledge of the free energy
of the magnetically ordered phase, which has not yet been obtained.

\section{Local stability of the partially ordered phase}
\label{sec:stability}

Even an analysis of the local stability of the partially ordered collinear
phase is not exactly straightforward. The standard
large-$S$ method of computing the magnon
contribution to the free energy fails because of the existence
of unstable modes with a negative stiffness at zero temperature. The
instability merely reflects the fact that the collinear
states are not a local minimum of energy (\ref{eq-H0}).  The instability
is avoided at a (sufficiently high) finite temperature: the free energy
of spin fluctuations contributes a positive term to the spin stiffness.
In this Section we analyze the local stability of the collinear phase.

\subsection{Unstable modes}

To analyze the stability of a collinear state, we express the energy
of the system in terms of transverse spin fluctuations
$\bm\sigma_i\perp\hat\mathbf n$. By substituting $\mathbf S_i
\approx S (1-\bm\sigma_i^2/2 S^2) \eta_i \hat \mathbf n + \bm
\sigma_i$ into Eq. (\ref{eq-H0}) we obtain a spin-wave Hamiltonian
in the harmonic approximation,
\begin{equation}
    \label{eq-H-sigma2}
    \mathcal{H}^{(2)}=E_L + (J_1-2 J_2)\sum_i\bm\sigma_i^2
    +\frac{1}{2}\sum_{i,j}J_{ij}\,\bm\sigma_i\cdot\bm\sigma_j,
\end{equation}
where $E_L$ is the energy of the layered state.  The Ising variables
$\{\eta_i\}$ specifying the direction of a spin are absent from the
harmonic Hamiltonian (\ref{eq-H-sigma2}).  They affect the dynamics
of the system through the canonical commutation relations for the
transverse components of the spins.

\begin{figure}
\includegraphics[scale=0.302]{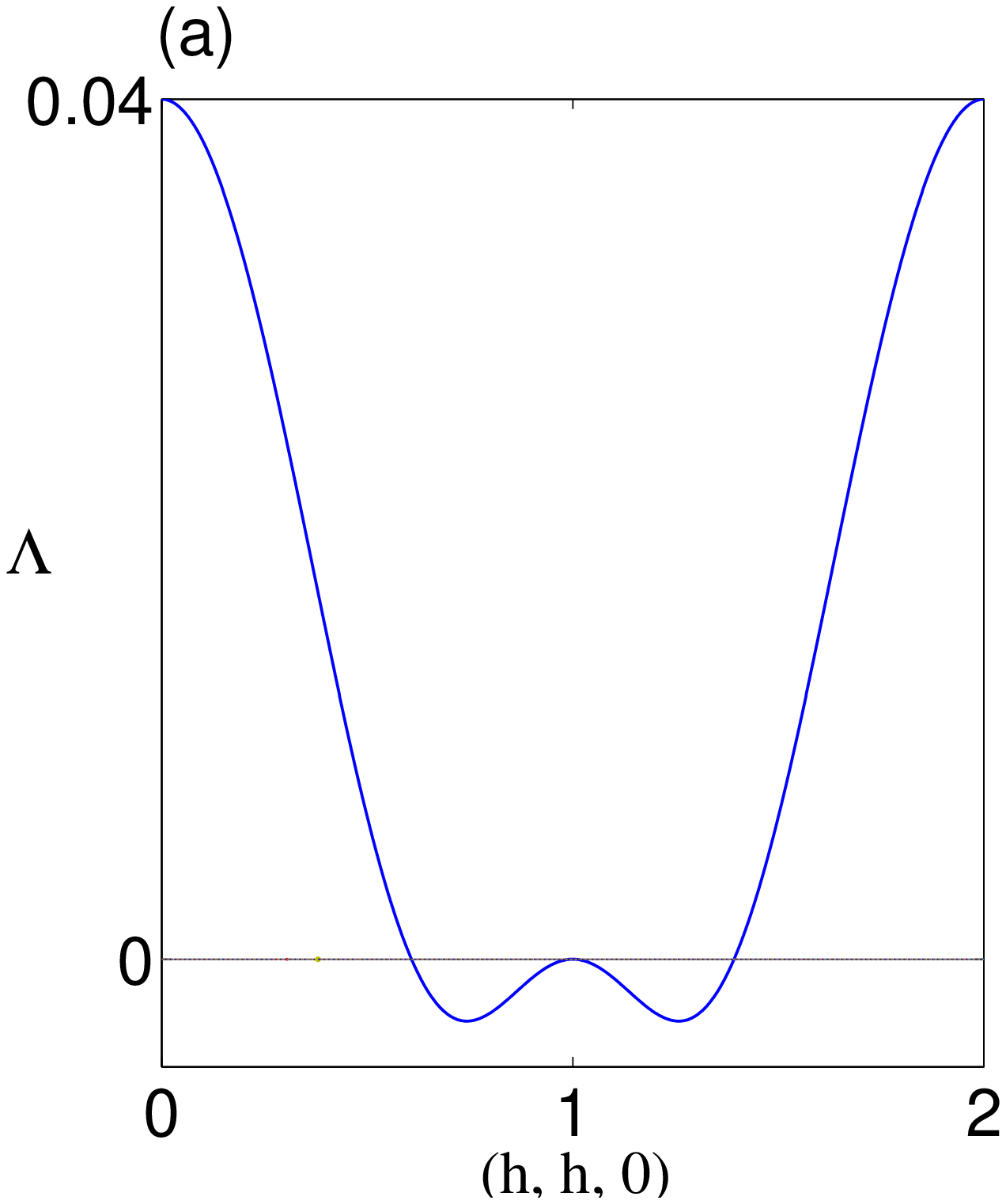}
\includegraphics[scale=0.302]{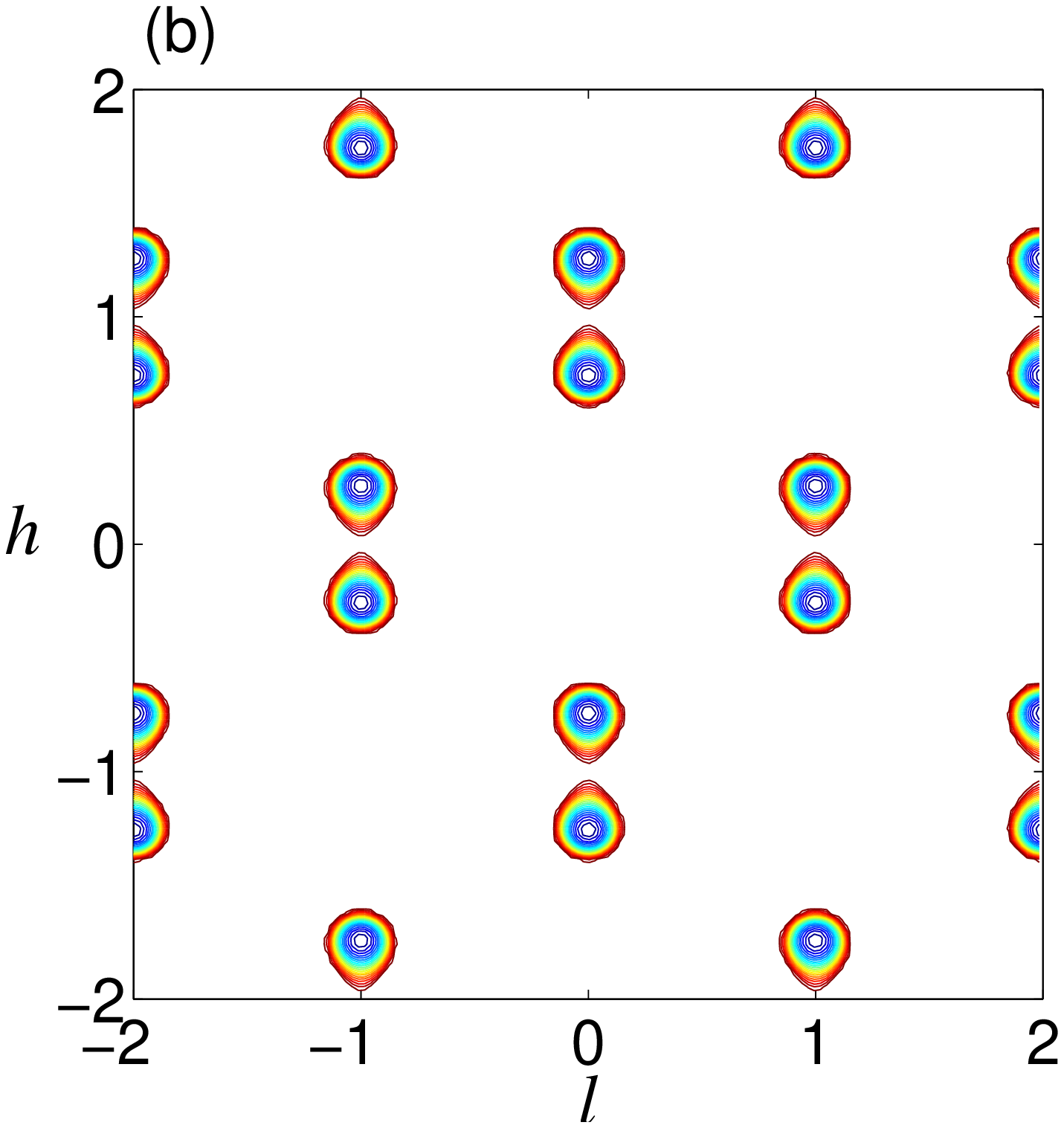}
\caption{\label{fig-unstable} (a) The energy dispersion of the spin-wave
band with unstable modes.
(b) Regions in momentum space $\mathbf q=2\pi(h,h,l)$
where the spectrum of energy fluctuations has negative eigenvalues
$\Lambda_{\mathbf q}^a$.  $J_1=1$, $J_2=-0.1$.}
\end{figure}

The quadratic form (\ref{eq-H-sigma2}) must be positive definite to
guarantee stability of the collinear state.  Its eigenvalues
$\Lambda$ are obtained by making the Fourier transform and then
diagonalizing a $4 \times 4$ matrix (the pyrochlore lattice is an
fcc with a basis of 4 sites):
\begin{equation}
    \Lambda^a_{\mathbf q} = (J_1-2 J_2)+\lambda^a_{\mathbf q},
\end{equation}
where $\lambda^a_{\mathbf q}$ are eigenvalues of $J_{mn}(\mathbf q)$
defined in Sec.~\ref{sec-high-frustration}. The dispersion has
degenerate zero modes along lines $\mathbf q=2\pi\{1,h,0\}$
corresponding to magnetic spirals along one of the three cubic axes.
These spirals belong to the degenerate manifold of non-collinear
layered states discussed in Appendix \ref{sec:layered-state}.
Furthermore, there are regions in momentum space with
$\Lambda_{\mathbf q} < 0$, as shown in Fig. \ref{fig-unstable}. The
most unstable modes are found at wavevectors $\mathbf
q^*=2\pi\{h^*,h^*,0\}$ with $h^*$ given by Eq.~(\ref{eq:h-j2}). For
small $J_2/J_1$, the lowest eigenvalue is
\[
    \frac{\Lambda_{\rm min}}{J_1}= (28-16\sqrt{3})\,\frac{J_2}{J_1}
    +\frac{32}{3}(56\sqrt{3}-97)\,\Bigl(\frac{J_2}{J_1}\Bigr)^2
    +\cdots.
\]
Since $\Lambda_{\rm min}<0$ for a ferromagnetic $J_2$, the collinear
ground states are unstable at zero temperature.

\subsection{Hartree-Fock calculation}

\begin{figure}
\includegraphics[scale=1.1]{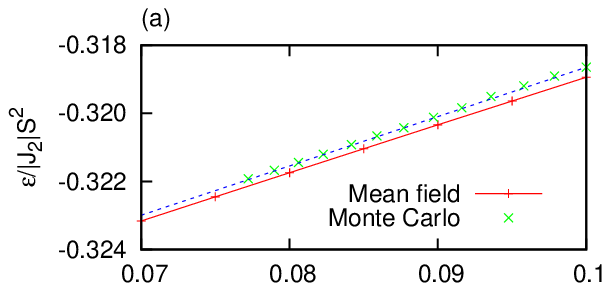}
\includegraphics[scale=1.1]{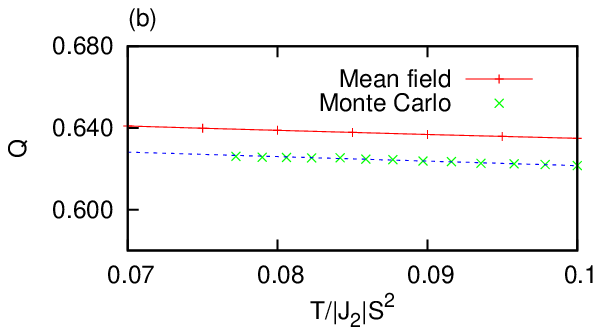}
\caption{\label{fig-comparison} (a) energy density $\varepsilon$ and
(b) nematic order parameter as a function of temperature obtained
using Monte Carlo simulations and a Hartree-Fock self-consistent
calculation. The calculation was done with $J_2=-0.01 J_1$. The
dashed line is a linear fit to the Monte Carlo data. Note that the
transition temperature obtained from Monte Carlo simulation is
$T_{c2}\approx 0.076\,|J_2| S^2$.}
\end{figure}

At finite temperatures the collinear layered states are stabilized
by thermal fluctuations. To demonstrate this, we go beyond the
harmonic term of the classical Holstein-Primakoff expansion and
consider the interactions between spin waves,\cite{hizi:2007jpcm}
\begin{eqnarray}
    \mathcal{H}^{(4)} = \frac{1}{8 S^2}\sum_{i,j} J_{ij}
    \Bigl[\eta_i\,\eta_j\,\bm{\sigma}_i^2\,\bm{\sigma}_j^2
    -\frac{1}{2}\bm{\sigma}_i\cdot\bm{\sigma}_j\,
    (\bm{\sigma}_i^2+\bm{\sigma}_j^2)\Bigr].
\end{eqnarray}
Since the system is unstable at the harmonic order,  a perturbation
expansion based on the quadratic Hamiltonian (\ref{eq-H-sigma2}) is
not possible.  Instead, following Hizi and
Henley,\cite{hizi:2007jpcm} we construct an effective (mean-field)
quadratic Hamiltonian
\begin{equation}
\mathcal{H}_\mathrm{MF} = \sum_{i,j} \tilde
H^{(2)}_{ij} \bm\sigma_i\cdot\bm\sigma_j
\label{eq:H-MF}
\end{equation}
that provides the best
approximation to $\mathcal H^{(2)} + \mathcal H^{(4)}$. To this end,
we use the standard mean-field recipe to decouple the quartic
Hamiltonian. We first write every possible pair of operators in
$\mathcal H^{(4)}$ in terms of its thermal average plus a
fluctuation term.  Dropping terms quartic in the fluctuations yields the
quadratic form (\ref{eq:H-MF}) with the following coefficients
$\tilde{H}^{(2)}_{ij}$:
\begin{eqnarray}
    \label{eq-HF}
       &(J_1-2 J_2)+\frac{1}{2 S^2}\,\sum_k\,J_{ik}\,
       (\eta_i\,\eta_k\,G_{kk}-G_{ik}) & (i=j),
     \nonumber\\
        &\frac{1}{2}\,J_{ij}\,\Bigl[1 + \frac{1}{S^2} \eta_i\,\eta_j\,G_{ij}
        -\frac{1}{2 S^2}\,(G_{ii}+G_{jj})\Bigr] & (i\neq j).
        \quad\quad
    \end{eqnarray}
Here $G_{ij}=\langle\sigma^x_i\sigma^x_j\rangle
=\langle\sigma^y_i\sigma^y_j\rangle$ is the correlation function of
spin fluctuations calculated self-consistently in the thermal
ensemble of the mean-field Hamiltonian (\ref{eq-HF})
\begin{equation}
    \label{eq-Gij}
    G_{ij} = \frac{\int\,D\bm{\sigma}\,\,\sigma^x_i\sigma^x_j
    \,e^{-\beta\mathcal{H}_{\rm MF}}}
    {\int\,D\bm{\sigma}\,e^{-\beta\mathcal{H}_{\rm MF}}}.
\end{equation}

\begin{figure}
\includegraphics[width=0.75\columnwidth]{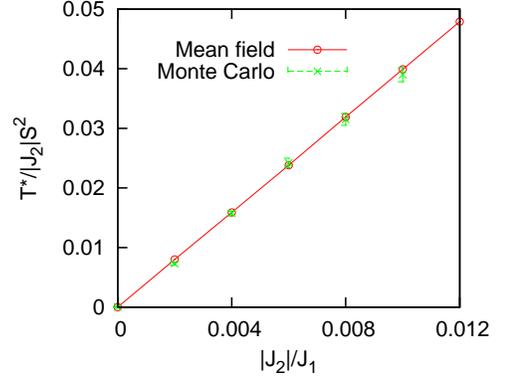}
\caption{\label{fig-t-bd} Stability boundary $T^*$ obtained using
the Hartree-Fock calculation and the Monte Carlo simulations. The error
bars shown for the Monte Carlo data are equal to the temperature step
$\Delta T$ used in the simulation.}
\end{figure}

Numerically, an iteration process is used to obtain the correlation
functions $G_{ij}$. After self-consistency is reached,
the energy of the magnet is given by
\begin{eqnarray}
    E_{\rm MF} &=& E_L + 2\sum_i (J_1-2 J_2)\,G_{ii}
    + \sum_{i,j} J_{ij}\,G_{ij} \\
    &+&\frac{1}{2 S^2}\sum_{i,j}J_{ij}\,\Bigl[\eta_i\eta_j
    (G_{ii} G_{jj}+G_{ij}^2)-G_{ij}(G_{ii}+G_{jj})\Bigr].
    \nonumber
\end{eqnarray}
Fig. \ref{fig-comparison} (a) shows the computed energy density as a
function of temperature. The result agrees very well with that
obtained from Monte Carlo simulations. Both the simulation and
calculation were done for $J_2=-0.01\,J_1$ on a pyrochlore lattice
with a size of $16\times 4^3$ spins and periodic boundary condition
on each side. The self-consistent method can also be used to compute
the nematic order parameter. For $\hat\mathbf n = + \hat \mathbf z$,
the tensor $\langle S_{\mu}S_{\nu} \rangle$ becomes diagonal with
elements $\langle S_x S_x \rangle = \langle S_y S_y \rangle = 2\,
\bar G$ and $\langle S_z S_z \rangle = 1-2\,\bar G$, where $\bar G =
\sum_i G_{ii}/N$. The nematic order parameter is then
\begin{equation}
    \label{eq-Q}
    Q = \frac{2}{3} - \frac{2}{N S^2}\sum_i G_{ii}.
\end{equation}
The result is shown in Fig. \ref{fig-comparison} (b) and the
agreement with that obtained from Monte Carlo simulation seems
satisfactory: the discrepancy between the two methods is less than
3\%. The nearly saturated nematic order parameter $Q$ observed in
Monte Carlo simulations implies $\bm\sigma^2\ll 1$, justifying the
Holstein-Primakoff expansion about the collinear state.

Below a certain temperature $T^*$ the energy spectrum of spin waves
acquires some negative eigenvalues and the collinear phase gives way
to the low-temperature ordered state. Since the transition is first
order, the $E-T$ diagram exhibits hysteresis. The thermodynamic
transition takes place at a temperature $T_{c2}>T^*$, at which the
collinear phase is still locally stable.

The dependence of $T^*/|J_2|$ on the ratio $|J_2|/J_1$ obtained from
the Hartree-Fock calculation is shown in Fig. \ref{fig-t-bd}. The
points collapse perfectly on a linear curve implying a scaling
relation $T^*\sim J_2^2/J_1$.  A numerical estimate of the stability
boundary $T^*(J_2)$, obtained as the lowest temperature at which the
intermediate phase was still observed in Monte Carlo runs, is also
plotted in Fig. \ref{fig-t-bd}; the result is in satisfactory
agreement with that of the mean-field calculation.

\subsection{Analytic results: red-and-green state}

An analytical derivation of the stability temperature $T^*\sim
J_2^2/J_1$ is difficult to obtain for the most general layered
state.  We have evaluated the stability for the simplest state of
this kind, where all of the layers have the same colors.  A state of
this sort (sublattice $A$ is red and sublattice $B$ is green) was
studied in Ref. \onlinecite{chern:060405}. This particular state has
a higher symmetry than a typical layered structure: the color
variables violate only the inversion symmetry exchanging the two
sublattices of tetrahedra.

\begin{figure}
\includegraphics[width=0.45\columnwidth]{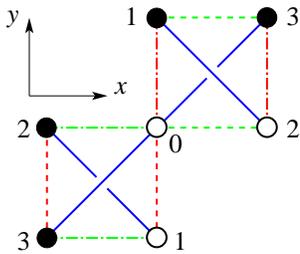}
\caption{\label{fig:J-dist} Renormalized nearest neighbor bonds of
the red-and-green state in the mean-field calculation. The
renormalized first-neighbor exchange constants: $J_1-K_1-K_2$
(dashed bonds), $J_1-K_1$ (dash-dotted bonds), and $J_1-K_2$ (solid
bonds).}
\end{figure}

In the mean-field Hamiltonian (\ref{eq-HF}), the main effects of the
quartic interaction $\mathcal{H}^{(4)}$ is to renormalize the
first-neighbor exchange $J_1$ to $J_{ij} = J_1 + \delta J_{ij}$,
which is now bond-dependent:
\begin{eqnarray}
    \label{eq:delta-J}
    \delta J_{ij} = -\frac{J_1}{2 S^2} (G_{ii}+G_{jj}-2\,\eta_i\eta_j G_{ij}).
\end{eqnarray}
Assuming that exchange renormalizations $\delta J_{ij}$ respect the
symmetries of the red-and-green state, we have 3 independent
variational parameters $\delta J_{01}$, $\delta J_{02}$, and $\delta
J_{03}$ (Fig. \ref{fig:J-dist}). If we further assume that the
correlations $G_{ij}$ are dominated by the pyrochlore zero modes,
the number of variational parameters reduces to 2. This is so
because zero modes satisfy $\sum_{i=0}^3\sigma_i = 0$, hence
$\langle \sigma_0 \sigma_1\rangle =
-\langle\sigma_0^2\rangle-\langle\sigma_0\sigma_2\rangle
-\langle\sigma_0\sigma_3\rangle$. It follows then that $\delta
J_{01} = \delta J_{02} + \delta J_{03}$.  We parameterize the
exchange renormalizations in terms of $K_1$ and $K_2$ such that
\begin{eqnarray}
    \label{eq:J-A}
    & & \delta J_{01} = \delta J_{23} = - K_1 - K_2, \nonumber \\
    & & \delta J_{02} = \delta J_{31} = - K_1, \\
    & & \delta J_{03} = \delta J_{12} = - K_2 \nonumber
\end{eqnarray}
on the red sublattice.

We then compute the spectrum and the eigenmodes of energy
fluctuations with the renormalized exchange interaction.  The two
zero-energy bands that were flat in the absence of $J_2$ and $K_i$
now acquire a dispersion; one becomes gapped ($\Lambda_\mathbf{q}^a$
is strictly positive), while the other has a vanishing energy at the
wavevector $\mathbf q_0 = 2\pi(0,0,1)$. This zero mode corresponds
to a global rotation of spins. Correlation functions are dominated
by fluctuations in the lowest band in the vicinity of $\mathbf q_0$.
For small $\mathbf k$, the energy eigenvalue is
\begin{eqnarray}
    \label{eq-disp1}
    \Lambda_\mathbf{q_0 + k} \approx \frac{1}{32}\bigl[
    2 K_1 k_{\perp}^2 + (8|J_2|+K_2) k_z^2\bigr],
\end{eqnarray}
where $k_{\perp}^2 = k_x^2+k_y^2$.

In order to obtain the correlations $G_{ij}$, we need first to
obtain the eigenmodes. To this end, we use an orthonormal basis of
the two zero modes of $J_1$ for given values of $\mathbf k$. We then
treat $K_i$ and $J_2$ as perturbations and use degenerate
perturbation theory to obtain the eigenmodes. To the lowest order in
$\mathbf k$, they are
\begin{eqnarray}
    u_0(\mathbf q_0 + \mathbf k) &=& -i/2 - (k_x - k_y + k_z)/16, \nonumber \\
    u_1(\mathbf q_0 + \mathbf k) &=& +1/2 -i(k_x + k_y + k_z)/16, \nonumber \\
    u_2(\mathbf q_0 + \mathbf k) &=& -1/2 -i(k_x + k_y - k_z)/16, \nonumber \\
    u_3(\mathbf q_0 + \mathbf k) &=& +i/2 - (k_x - k_y - k_z)/16.
\end{eqnarray}
As can be easily checked, the total spin of a tetrahedron
$\sum_m\sigma_m = \sum_m u_m e^{i(\mathbf q_0+\mathbf k)\cdot\mathbf
r_m} = 0$ at this order of $k$. The spin correlation function is
\begin{eqnarray}
    \label{eq-Gmn}
    G_{mn} = \frac{1}{N'}\sum_{\mathbf q}
             \frac{T}{\Lambda_{\mathbf q}}
    u^*_m(\mathbf q)\,u_n(\mathbf q)\,
    e^{i(\mathbf q)\cdot(\mathbf r_m-\mathbf r_n)},
\end{eqnarray}
where $N' = N/4$ is the number of unit cells, and $m$, $n$ are
sublattice indices. By expanding to the second order of $k$ and
using (\ref{eq:delta-J}), we obtain the following self-consistency
equations for $K_1$ and $K_2$
\begin{eqnarray}
    \label{eq:K1}
    \frac{J_1 T}{4N' S^2}\sum_{\mathbf k}
    \frac{k_z^2}{2 K_1\,k_{\perp}^2+(8|J_2|+K_2)k_z^2} = K_1, \\
    \label{eq:K2}
    \frac{J_1 T}{2N' S^2}\sum_{\mathbf k}
    \frac{k_{\perp}^2}{2K_1\,k_{\perp}^2+(8|J_2|+K_2)k_z^2} = K_2,
\end{eqnarray}

Although these equations can be solved numerically, we are
interested in an approximate solution of $K_1$ and $K_2$ in the
low-temperature regime, $T\ll |J_2| S^2$. Since the effective spin
stiffness $K$ is generated by thermal fluctuations, they are
expected to be small compared to $J_2$. To the lowest order we
neglect $K_1$ and $K_2$ in Eq.~(\ref{eq:K1}) and obtain
\begin{eqnarray}
    K_1 \approx \frac{J_1 T}{32\, |J_2| S^2}.
\end{eqnarray}
On the other hand, because the integral for $K_2$ is divergent as
$K_1\to 0$, we must keep $K_1$ in Eq.~(\ref{eq:K2}). Substituting
the result for $K_1$ into Eq.~(\ref{eq:K2}), we obtain
\begin{eqnarray}
    K_2 \approx \frac{\pi}{3\sqrt{2} S}\sqrt{J_1 T}
\end{eqnarray}
to the lowest order in $T$.

\begin{figure}
\includegraphics[width=0.65\columnwidth]{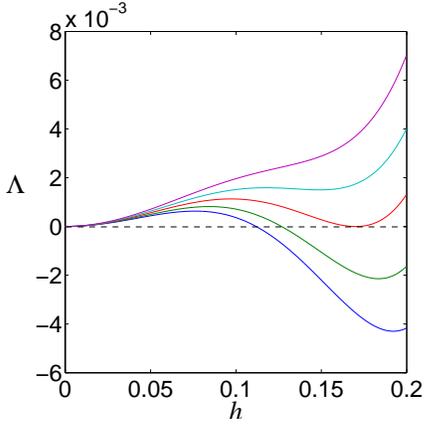}
\caption{\label{fig-trans-n} Variation of spin-wave energy $\Lambda$
in unit of $|J_2|$ along the $\mathbf q = 2\pi(h,h,1)$ line. The
calculation was done with a $J_2 = -0.01 J_1$. The curves correspond
to temperatures $T/|J_2|=$ 0.018, 0.0165, 0.01526, 0.0145, 0.013
(from top to bottom) $T^* = 0.01526 |J_2| S^2$ corresponding to the
temperature where the ${\bf q}=0$ mode becomes unstable.}
\end{figure}

These results provide a glimpse into the physics of the transition
between the intermediate and low-temperature phases. Fig.
\ref{fig-trans-n} shows the renormalized dispersion
$\Lambda_\mathbf{q_0 + k}$ (\ref{eq-disp1}) along the line $\mathbf
q=2\pi(h,h,1)$ at various temperatures. As the temperature
decreases, a dip of the dispersion curve starts to develop at
$h\approx 0.2$. Eventually this local minimum touches zero at the
critical temperature $T_{c2}$; below $T^*$ the collinear state is
unstable: it decays by emitting spin waves with $\mathbf q \approx
2\pi(1/4,1/4,1)$, which is related to $2\pi(3/4, 3/4, 0)$ by a
reciprocal lattice vector.

It should be noted that the scenario displayed in Fig.
\ref{fig-trans-n} is only a qualitative description of the real
transition. Our self-consistent treatment only takes into account
spin waves close to the $\mathbf q_0=2\pi(0,0,1)$ Goldstone mode.
This is valid at temperatures well above $T^*$ since these spin
waves are the lowest-energy excitations of the magnet. However, as
$T\to T_{c2}$, spin waves with wavevectors $\mathbf q \approx
2\pi(3/4,3/4,0)$ become soft and should also be included in a
self-consistent calculation.  Additionally, we have studied the
energy of spin waves as a proxy for the instability, whereas the
proper calculation at a finite temperature should involve the free
energy. We do this next.

\subsection{Stability boundary: red-and-green state}

We now provide an estimate of the stability temperature $T^*$ by
computing the magnon contribution to the system free energy. An
expression (\ref{eq-F-unstable}) for the change of free energy
associated with an unstable mode is derived in Appendix
\ref{sec-fe-unstable}. Here we apply the result to the red-and-green
state. We consider the most dangerous modes, namely those with
wavevectors near $\mathbf q^* = 2\pi\{h^*,h^*,0\}$ where $h^*\approx
3/4$. In the presence of such an unstable mode with amplitude $\phi$
superimposed on the red-and-green state, the free energy changes by
an amount given by
\begin{equation}
    \label{eq-DF}
    \Delta F = \Bigl(\Lambda^* S^2 + \sum_{mn}
    G_{mn}\,\Delta_{nm}\Bigr)\phi^2,
\end{equation}
where the correlation function $G_{mn}$ is given by (\ref{eq-Gmn}),
and $\Delta_{mn}$ is the perturbation to the mean-field Hamiltonian
$\mathcal{H}_{\rm MF}$ caused by the unstable mode. In our case, the
real-space eigenvector of the unstable mode with $\mathbf q^*
=2\pi(h^*,h^*,0)$ is
\begin{equation}
    \mathbf m_{n}(\mathbf r) = U^*_{n}\,
    \bigl[\hat\mathbf x\cos(\mathbf q^*\cdot\mathbf r)
    +\hat\mathbf y\sin(\mathbf q^*\cdot\mathbf r)\bigr],
\end{equation}
where the corresponding momentum-space eigenvector for $\mathbf q^*$
is
\begin{equation}
    \label{eq-eigenU}
    \mathbf U^*
    = (\cos\theta,\,-\sin\theta,\,-\sin\theta,\,\cos\theta)/\sqrt{2},
\end{equation}
with $\theta \approx 0.27 \pi$ and weakly dependent on $J_2$. We
write the energy of the unstable mode as $\Lambda^* =
-\gamma\,|J_2|$, where $\gamma \approx 0.2$ is a dimensionless
number. The change of free energy is then
\begin{equation}
    \Delta F/\phi^2 = -\gamma|J_2| S^2 + \frac{J_1 T}{4 N'}
    \sum_{\mathbf k} \frac{\Delta_\mathbf k}
    {2 K_1\,k_{\perp}^2+(8|J_2|+K_2)k_z^2},
\end{equation}
where
\begin{equation}
    \Delta_{\mathbf k} =\sum_{m,n} \Delta_{mn}
    u^*_n(\mathbf k)u_m(\mathbf k)e^{i(\mathbf q_0+\mathbf k)
    \cdot(\mathbf r_m-\mathbf r_n)}.
\end{equation}
Since in most cases $\Delta_{\mathbf k}\sim \Delta
+\mathcal{O}(k^2)$ for $h^*=1/4$, we neglect the $\mathbf k$
dependence of $\Delta_{\mathbf k}$ in the following as a lowest
order approximation. With the aid of Eqs.~(\ref{eq:K1}) and
(\ref{eq:K2}), the integral evaluates to
\begin{equation}
    \frac{\sqrt{2}\,\Delta}{16\pi}\,\sqrt{J_1 T},
\end{equation}
The condition $\Delta F = 0$ thus gives an estimate of the stability
temperature
\begin{equation}
    \label{eq-T-star}
    T^*=\Bigl(\frac{16\pi \gamma S}{\sqrt{2}\Delta}\Bigr)^2
    \,\frac{J_2^2}{J_1}.
\end{equation}
This expression overestimates (by a factor of about 10) the
stability temperature compared with numerical results. However, as
mentioned previously, the discrepancy is due to the fact that we
neglect contributions from the unstable modes themselves when
approaching the transition temperature. Those modes with wavevector
centered about the 12 unstable $\mathbf q^*=2\pi\{h^*,h^*,0\}$
become extremely soft as $T\to T^*$ and should be included in the
calculation in a self-consistent way. Nevertheless,
(\ref{eq-T-star}) provides an upper bound of the stability boundary
and gives a scaling relation consistent with the numerical data.

\section{Discussion}
\label{sec:discussion}

We have studied the classical Heisenberg antiferromagnet on the
pyrochlore lattice with first and second-neighbor exchange
interactions.  Ferromagnetic second-neighbor exchange $J_2<0$ is
frustrated and lifts the vast degeneracy of the nearest-neighbor
model only partially, setting stage for a nontrivial phase diagram
in the $(J_2, T)$ plane.  We have used a combination of Monte Carlo
simulations and analytical calculations to characterize the phases
of this model. In our opinion, the low-temperature phase, discussed
previously by Tsuneishi {\it et al.},\cite{tsuneishi:2007jpcm}  is
the incommensurate, and likely non-collinear, ordered phase
predicted earlier by Reimers {\it et al}.\cite{reimers:1991prb}  A
full characterization of its magnetic order remains to be done, and
its fate in the presence of strong quantum fluctuations is an
interesting topic for future study.

Our simulations have uncovered the existence of another,
partially-ordered phase at intermediate temperatures for a weak
enough $|J_2|$.  In the intermediate phase, the spins are on average
collinear, which is manifested by a nonzero nematic order parameter.
The order is fully characterized by a combination of a global
nematic director $\hat\mathbf n$ and a 3-state Potts variable
(color) on every tetrahedron indicating the location of frustrated
bonds (Fig.~\ref{fig:rgb}).  The second-neighbor interaction $J_2<0$
acts like an antiferromagnetic Potts coupling forcing unlike colors
on neighboring tetrahedra.

The color structure of this phase resembles the ordered state with
broken sublattice symmetry (BSS)  of the antiferromagnetic Potts
model: \cite{grest:1981prl} one sublattice of tetrahedra is
dominated by one color (say, blue) while the other exhibits a
mixture of the remaining two colors (red and green). However, unlike
in the BSS state, the two colors on the second sublattice are not
distributed in a completely random way: they form uniform layers in
the plane associated with the colors (in this case, $xy$).  The
colors of individual layers appear to be random, hence
\textit{partial} order.

The partial order can be described by an individual $Z_2$ variable
$\sigma_i$ for each such layer---in addition to a global direction
of the spins $\hat\mathbf n$ and the color of the other sublattice.
States with different sets of $\{\sigma_i\}$ are local minima of the
free energy.  Accessing one such minimum from another by means of a
uniform rotation of spins within one layer of tetrahedra requires
climbing over a free-energy barrier that grows as the number of
spins in that layer and thus becomes impossible in the thermodynamic
limit. A more plausible route to changing the color of a layer is by
nucleating a bubble of the opposite $\sigma_i$, which will grow if
the new state has a lower free energy once the bubble is large enough for the
gain in bulk energy to outweigh the cost in interface energy. Since the
distinct layered states are not related by symmetry, their free energies are
generally different and the nucleation route may well lead to a
selection within this set of states. Since such nucleation can go along with
large energy barriers, it can be tricky to observe
\cite{leggett:1984prl,schiffer:1992prl}, and indeed we have not found it in our
simulations.

It is worth stressing that the ideal collinear states do not
minimize the exchange energy---either globally or locally.  They owe
their stability to thermal fluctuations, which effectively
renormalize exchange couplings and turn these spin configurations
into minima of the free energy.  As the temperature falls, the
couplings return to their bare values and the collinear states
become locally unstable at a temperature $T^* = \mathcal
O(J_2^2/J_1)$, in agreement with our Monte Carlo simulations.  The
most unstable spin-wave mode has approximately the same wavenumber
as the low-temperature incommensurate magnetic order.  The simulated
phase transition is strongly discontinuous.

Simulations on the high-temperature side show that the intermediate
phase persists up to a temperature $\mathcal O(J_2)$.  A
discontinuous phase transition takes it into the paramagnetic phase.
The presence of strong local spin correlations in the paramagnetic
phase means that the effect of third-neighbor couplings $J_3$ (but
not of $J_3'$, see Fig.~\ref{fig:pyrochlore}) is equivalent, up to a
change of sign, to that of the second-neighbor coupling, at least to
the first order.  Therefore we expect that the state of our system
depends on these couplings mostly through their difference
$J_2-J_3$.  If correct, this observation would extend the results of
our study to a broader class of pyrochlore antiferromagnet with both
$J_2$ and $J_3$ present.

\section*{Acknowledgments}

It is a pleasure to thank J. Chalker and G. Jackeli for useful
discussions. This work was supported in part by the NSF under Grant
No. DMR-0348679 and by Research Corporation.

\appendix

\section{Equivalence of $J_2$ and $-J_3$ in the strongly correlated paramagnet}
\label{app:equiv}

At temperatures $T \ll J_1 S^2$ spins on every tetrahedron
approximately satisfy the
constraint 
\begin{equation}
\sum_{i=0}^3 \mathbf S_i = 0.
\label{eq:constraint}
\end{equation}
Consider the effective magnetic field on the site labeled $3'$ in
Fig.~\ref{fig:pyrochlore}:
\begin{equation}
\mathbf H_{3'} = -\partial H/\partial \mathbf S_{3'}
= -J_1 \mathbf S_0 - J_2 (\mathbf S_1 + \mathbf S_2) + \ldots
\label{eq:Heff-1}
\end{equation}
where we have explicitly written out the contributions from the
spins of the adjacent tetrahedron 0123.  Let us now turn off the
second-neighbor exchange, $J_2=0$, and turn on the third-neighbor
coupling $J_3$ (Fig.~\ref{fig:pyrochlore}).  Doing so changes the
effective field to
\begin{equation}
-J_1 \mathbf S_0 - J_3 \mathbf S_3
\approx -(J_1 - J_3) \mathbf S_0 + J_3(\mathbf S_1 + \mathbf S_2),
\label{eq:Heff-2}
\end{equation}
where we used the constraint (\ref{eq:constraint}).  In this
setting, a comparison of Eqs.~(\ref{eq:Heff-1}) and
(\ref{eq:Heff-2}) shows that adding a third-neighbor coupling $J_3$
is indeed energetically equivalent to the second-neighbor exchange
of the same magnitude and opposite sign.
This result does not extend to the excited states, which violate
(\ref{eq:constraint}), so that the physics of fluctuations need
not be simply related.

\section{Constraint on colors (bond variables)}
\label{app:constraint}

\begin{figure}
\includegraphics[width=0.8\columnwidth]{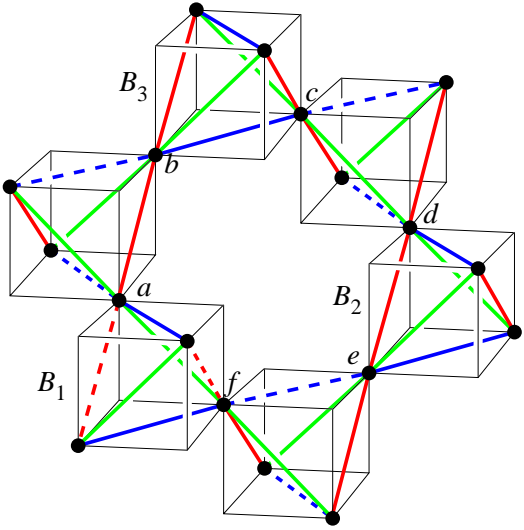}
\caption{\label{fig-hex} A fragment of a collinear state.
Frustrated bonds are shown as colored dashed lines.}
\end{figure}

Consider the hexagonal loop $abcdef$ shown in Fig. \ref{fig-hex}.
Suppose that tetrahedra of sublattice $A$ are in the blue state and
that one tetrahedron $B_1$ of the other sublattice is red. Then it
can be seen that tetrahedron $B_2$, which has the same $z$
coordinate, must also be red. This can be proved as follows. In the
collinear state, spins can be represented by a Ising variable, i.e.
$\mathbf S_i = S \eta_i\hat\mathbf n$ and $\hat\mathbf n$ is an
arbitrary unit vector. Obviously, the product of the six bond
variables $\eta_i\eta_j$ on the hexagon loop is $+1$, i.e.
\begin{equation}
    (\eta_a\eta_b)(\eta_b\eta_c)\cdots(\eta_e\eta_f)(\eta_f\eta_a) = +1.
\end{equation}
Among the six bonds, sublattice $A$ contributes two
antiferromagnetic and one ferromagnetic bond; this makes its total
contribution $+1$. Therefore the product of the three bonds on
sublattice $B$ must be $+1$ as well. We know that $\eta_f\eta_a =
+1$ ($B_1$ is red) and $\eta_b\eta_c = -1$ ($B_3$ is not blue).
Hence $\eta_d\eta_e = -1$, which means that $B_2$ is not green.
Since $B_2$ is not blue, it must be red.

Thus, if sublattice $A$ is blue, the above proof shows that
sublattice $B$ has uniform color green or red in each layer $z$ =
const. However, the color of individual $B$ layers are random. This
is similar to the BSS phase of antiferromagnetic Potts model where
individual sites on one sublattice have random colors.

\section{Magnetic structure of the layered state}
\label{sec:layered-state}

At the mean-field level, the collinear BSS-like states of the
intermediate phase are degenerate with a larger class of layered
state with non-collinear spins in general. Here we describe the
magnetic structure of the general layered state.

At temperatures well below the Curie-Weiss constant, the magnetic
state of a tetrahedron is determined by three staggered
magnetizations $\mathbf L_i$, where $\mathbf L_1 = (\mathbf S_0 +
\mathbf S_1 - \mathbf S_2-\mathbf S_3)/4$, and so on.
\cite{chern:060405} Here we choose to specify the N\'eel vectors of
layers belonging to sublattice $A$. Because each spin is shared by
two tetrahedra from different sublattices, the magnetic state of
tetrahedra of sublattice $B$ is encoded in the staggered
magnetizations of the four surrounding tetrahedra of sublattice $A$.

\begin{figure}
\includegraphics[width=0.8\columnwidth]{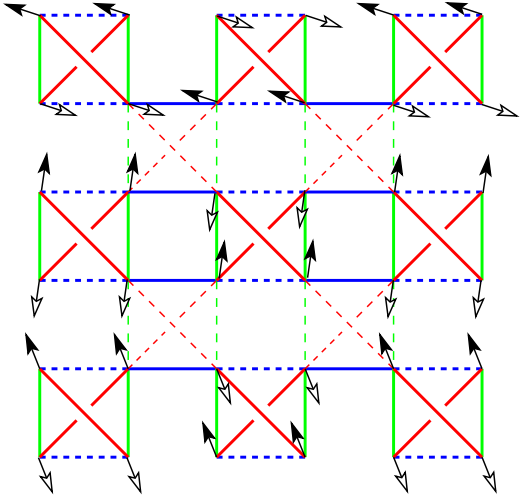}
\caption{\label{fig:layer} Non-collinear layered state projected
along the $x$-axis.
The normal of the layers are parallel to the $z$ axis.
Frustrated bonds are shown as dashed lines.}
\end{figure}

Choosing the normal of the layers to be the $z$ axis, the
staggered magnetizations of a tetrahedron with inplane
coordinate vector $\mathbf r_{\perp}$ in the $k$-th layer are
\begin{eqnarray}
    \mathbf L_1 = \mathbf L_2 = 0,\quad
    \mathbf L_3 = S\hat\mathbf n_k\,
    e^{i\mathbf q_{\perp}\cdot\mathbf r_{\perp}},
\end{eqnarray}
where $\mathbf q_{\perp} = 2\pi(1,0)$ or $2\pi(0,1)$ which are
equivalent with respect to the two-dimensional (2D) square lattice
of tetrahedra with the same $z$ coordinate, and $\hat\mathbf n_k$ is
an arbitrary unit vector. Fig. \ref{fig:layer} shows an example of
the 2D N\'eel order. Once the spin order within the layers is
specified, the magnetic structure of a general layered state is
described by a sequence of the unit vectors $\hat\mathbf n_k$.

The bond order of the layered state is similar to the BSS state of
three-state Potts model. In the example given above, sublattice $A$
is in the collinear blue state while tetrahedra in sublattice $B$ in
general have coplanar spins; their bond order is determined by the
N\'eel vectors of the two $A$ layers enclosing it:
\begin{equation}
    \mathbf f_B = \frac{4
    S^2}{\sqrt{3}}\Bigl(\frac{1}{2}\,,\,\frac{\sqrt{3}}{2}\,\hat\mathbf n_k
    \cdot\hat\mathbf n_{k+1}\Bigr).
\end{equation}
For arbitrary $\hat\mathbf n_k$, the bond vector $\mathbf f_B$
spreads uniformly on the edge of the triangle domain which connects
the two vertices corresponding to the red and green states; the
average color is again yellow.

The energy of a layered state is independent of the direction
$\hat\mathbf n_k$ of spins in the individual layers:
\begin{equation}
    \label{eq:EL}
    E_L = -N (J_1 - 2 J_2) S^2.
\end{equation}
This energy corresponds to the extrapolated zero-temperature energy
density $\varepsilon_L = -|J_2|/3$ of the nematic phase in Fig.
\ref{fig-transition}. Although all layered states are degenerate at
the mean-field level, thermal fluctuations apparently prefer the
collinear ones as shown by the Monte Carlo simulations.

In the collinear layered states, a common direction $\hat\mathbf n$
is selected and the unit vectors $\hat\mathbf n_k \to
\sigma_k\,\hat\mathbf n$ with the Ising variable $\sigma_k = \pm 1$.
The magnetic structure of a layered state is then specified by a
sequence of Ising variables:
$\{\sigma_1,\sigma_2,\cdots,\sigma_k,\cdots\}$. For a pyrochlore
lattice with 8 layers of tetrahedra in each direction, there are 17
distinct layered states that are not related to each other by
translations and inversions of the Ising variables (Table I).
However, some of these states may be related by other symmetries of
the lattice. For example, both states 1 and 17 represent the
red-and-green state.

\begin{table}
\label{tab:collinear}
\begin{tabular}{|l|c|l|c|}
 \hline
 No. & Ising sequence & 9 & $--++-++\,+$ \\
 \hline
 1 & $+++++++\,+$ & 10\,\,\, & $-+-+-++\,+$ \\
 \hline
 2 & $-++++++\,+$ & 11 & $-++-++-\,+$ \\
 \hline
 3 & $--+++++\,+$ & 12 & $----+++\,+$ \\
 \hline
 4 & $-+-++++\,+$ & 13 & $---+-++\,+$ \\
 \hline
 5 & $-++-+++\,+$ & 14 & $---++-+\,+$ \\
 \hline
 6 & $-+++-++\,+$ & 15 & $--+-+-+\,+$ \\
 \hline
 7 & $---++++\,+$ & 16 & $--+-++-\,+$ \\
 \hline
 8 & $--+-+++\,+$ & 17 & $-+-+-+-\,+$ \\
\hline
\end{tabular}
\caption{Ising sequences of the 17 distinct layered states
for a pyrochlore lattice with 8 layers subject to periodic
boundary conditions.}
\end{table}

\section{Free energy of the unstable mode}
\label{sec-fe-unstable}

Below we derive a Holstein-Primakoff Hamiltonian for spin waves in a
state with nearly collinear spins and compute the free energy to
test the local stability of the collinear state.  We focus
specifically on the previously identified unstable modes.

In a general state with noncollinear spins ${\bf S}_i = S \hat{\bf
n}_i$, we introduce a local reference frame defined by three
orthonormal vectors: $\hat{\bf e}^x_i$, $\hat{\bf e}^y_i$, and
$\hat{\bf n}_i$. A small deviation from this state can then be
expressed using the Holstein-Primakoff expansion:
\begin{equation}
    {\bf S}_i = S \Bigl(1-\frac{\bm{\sigma}_i^2}{2 S^2}\Bigr)\,\hat{\bf n}_i
    +\sum_{\alpha=x,y} \sigma^\alpha_i\, \hat{\bf e}^\alpha_i
    + \mathcal{O}(\sigma^3),
\end{equation}
Here $\bm\sigma_i = (\sigma^x_i, \sigma^y_i)$ whose components
denote fluctuation along the two orthogonal local axes.
The exchange Hamiltonian then becomes
\begin{equation}
    \mathcal{H} =\sum_{i,j} J_{ij}\ \hat{\bf n}_i\cdot\hat{\bf n}_j
    +\sum_{i,j}\sum_{\alpha,\beta} H^{\alpha\beta}_{ij}\,\sigma^\alpha_i\;\sigma^\beta_j.
\end{equation}
with
\begin{equation}
    \label{eq-H-noncollinear}
    H^{\alpha\beta}_{ij}=\left\{
    \begin{array}{ll}
        -\frac{1}{2}\sum_k J_{ik}\,\hat{\bf n}_i\cdot\hat{\bf n}_k\,
        \delta^{\alpha\beta} & (i=j), \\
        \frac{1}{2}\,J_{ij}\,\hat{\bf e}^\alpha_i\cdot\hat{\bf e}^\beta_j & (i\neq j).
    \end{array}\right.
\end{equation}
For collinear states in the nematic phase, $\hat{\bf
n}_i=\eta_i\hat{\bf z}$, and $\hat{\bf e}^x_i=\hat{\bf x}$,
$\hat{\bf e}^y_i=\hat{\bf y}$, Eq. (\ref{eq-H-noncollinear})
reproduces the harmonic Hamiltonian (\ref{eq-H-sigma2}).

Next we compute the increase in the free energy resulting from a
small deviation from a collinear state in the direction of an
unstable mode of the bare Hamiltonian (\ref{eq-H-sigma2}). Let
$\phi$ be the amplitude of the unstable mode and $\{{\bf m}_i\}$ the
corresponding (normalized) real-space eigenvector. The deformed spin
configuration is
\begin{equation}
    {\bf S}_i = S\hat\mathbf n_i = S\hat{\bf z} \, \eta_i(1-\phi^2{\bf m}_i^2/2)
    + S \phi\, {\bf m}_i .
\end{equation}
Given this local spin axis $\hat\mathbf n_i$, there is arbitrariness
in the choice of the other two unit vectors $\hat{\bf e}^\alpha_i$.
In order to apply the perturbation method, we choose:
\begin{eqnarray}
    \hat{\bf e}^x_i = \hat{\bf x}\,\Bigl[1-\phi^2\frac{(m^x_i)^2}{2}\Bigr]
    -\hat{\bf y}\,\phi^2\frac{m^x_i m^y_i}{2}
    - \hat{\bf z}\,\eta_i\,\phi\, m^x_i, \nonumber \\
    \hat{\bf e}^y_i = -\hat{\bf x}\,\phi^2\frac{m^x_i m^y_i}{2}
    + \hat{\bf y}\,\Bigl[1-\phi^2\frac{(m^y_i)^2}{2}\Bigr]
    -\hat{\bf z}\,\eta_i\,\phi\, m^y_i.
\end{eqnarray}
Substituting these expressions into Eq. (\ref{eq-H-noncollinear}),
we obtain $H^{\alpha\beta}_{ij} = \delta^{\alpha\beta}\,H^{(2)}_{ij}
+\phi^2 \Delta^{\alpha\beta}_{ij}$, where $H^{(2)}_{ij}$ is the
magnon Hamiltonian of the collinear state and the perturbation
$\Delta^{\alpha\beta}_{ij}$ is given by
\begin{equation}
    \label{eq-Delta1}
    \Delta^{\alpha\beta}_{ij} = \left\{\begin{array}{ll}
    -\frac{1}{2}\sum_k J_{ik}\bigl[{\bf m}_i\cdot{\bf m}_k
    -\frac{1}{2}\eta_i\eta_k({\bf m}_i^2+{\bf m}_k^2)\bigr]\delta^{\alpha\beta}, & (i=j) \\
    \frac{1}{2}\, J_{ij} \bigl[\eta_i\,\eta_j\, m^\alpha_i\, m^\beta_j
    -\frac{1}{2} (m^\alpha_i m^\beta_i + m^\alpha_j m^\beta_j) \bigr], & (i\neq j)
    \end{array}\right.
\end{equation}
Since the bare harmonic Hamiltonian contains unstable modes as
discussed in Section \ref{sec:stability}, we replace $H^{(2)}_{ij}$
by the one renormalized by spin-wave interactions, given in Eq.
(\ref{eq-HF}). We may then approximate the free energy of the system
as
\begin{eqnarray}
    \label{eq:Z1}
    e^{-\beta F} &\approx& e^{-\beta\, (E_L + \Lambda^* S^2 \phi^2)} \nonumber\\
    & &\;\times \int' D\bm{\sigma}\,\, e^{-\beta\sum_{i,j}
    \bigl[ \tilde H^{(2)}_{ij}\bm{\sigma}_i\cdot\bm{\sigma}_j
    + \phi^2\;\Delta^{\alpha\beta}_{ij}
    \sigma^\alpha_i\,\sigma^\beta_j\bigr]} \nonumber \\
    &=& \tilde Z\;e^{-\beta\, (E_L + \Lambda^* S^2 \phi^2)}\;
    \Bigl\langle e^{-\beta\phi^2 \sum_{i,j} \Delta^{\alpha\beta}_{ij}\,\;
    \sigma^\alpha_i\,\sigma^\beta_j}\Bigr\rangle \nonumber \\
    &\approx& \tilde Z\;e^{-\beta\, (E_L + \Lambda^* S^2 \phi^2)}\;\;
    e^{-\beta\phi^2 \sum_{i,j} \Delta^{\alpha\beta}_{ij}
    \langle\sigma^\alpha_i\,\sigma^\beta_j \rangle}. \nonumber \\
\end{eqnarray}
Here $E_L$ is energy of the layered state, $\tilde Z$ is the
partition function of the renormalized Hamiltonian
$\tilde\mathcal{H}^{(2)}$, $\langle\cdots\rangle$ means Boltzmann
averaging with respect to the Hamiltonian $\tilde\mathcal{H}^{(2)}$,
and $\Lambda^* < 0$ is the bare energy of the unstable mode $\mathbf
m_i$. The prime in the integral indicates that we only integrate out
the low-energy magnons close to the Goldstone mode of the collinear
state.

Upon expanding the fluctuations in terms of spin-wave eigenvectors
$\sigma^\alpha_i = \sum_n \xi^\alpha_n\,u_{n,i}$, we obtain the spin
correlation:
\begin{eqnarray}
    \langle\sigma^\alpha_i \sigma^\beta_j\rangle
    &= \delta^{\alpha\beta} \sum_n\!' \langle |\xi_n|^2 \rangle\; u_{n,i}^*\,u_{n,j} \nonumber \\
    = \delta^{\alpha\beta}& \sum_n\!' \frac{T}{\Lambda_n}\; u_{n,i}^*\,u_{n,j}
    = \delta^{\alpha\beta}\,G_{ij}.
\end{eqnarray}
Here $\Lambda_n$ is the energy of the $n$-th eigenmode of the
renormalized Hamiltonian $\tilde H^{(2)}_{ij}$. Substituting this
result back into Eq. (\ref{eq:Z1}) yields the free energy
(\ref{eq-DF}) associated with the unstable mode $\phi$,
\begin{eqnarray}
    \label{eq-F-unstable}
    F \approx \mbox{const} + \Bigl(\Lambda^* S^2 +
    \sum_{i,j} G_{ij}\Delta_{ji}\Bigr)\phi^2
    + \mathcal{O}(\phi^4),
\end{eqnarray}
where $\Delta_{ij} = \sum_\alpha \Delta^{\alpha\alpha}_{ij}$.

\end{document}